%% file: main.tex
\documentclass[acmsmall]{acmart}

\usepackage{xcolor}
\definecolor{ourblue}{rgb}{.318,.529,.765}
\definecolor{ourgreen}{rgb}{.412,.706,.329}
\definecolor{ourorange}{rgb}{1,.533,.067}
\usepackage{tkz-kiviat} 
\usetikzlibrary{arrows}

\usepackage[english]{babel}
\usepackage[utf8x]{inputenc}
\usepackage[T1]{fontenc}

\usepackage[disable]{todonotes}
\usepackage{graphicx}
\usepackage{hhline}

\usepackage{caption}
\usepackage[english]{babel}
\usepackage[utf8x]{inputenc}
\usepackage[T1]{fontenc}

\usetikzlibrary{shapes,snakes}
\newcommand*\circled[1]{\tikz[baseline=(char.base)]{
		\node[shape=circle,draw,inner sep=0.8pt] (char) {#1};}}

\usepackage{multirow,tabularx,environ,colortbl,ltablex,ragged2e}

\newcolumntype{s}{>{\RaggedRight\hsize=.7\hsize}X}
\newcolumntype{j}{>{\RaggedRight\hsize=.7\hsize}X}
\newcolumntype{b}{>{\RaggedRight\hsize=1.3\hsize}X}
\newcolumntype{k}{>{\RaggedRight\hsize=1.3\hsize}X}

\usepackage[absolute,showboxes]{textpos}

\usepackage[framemethod=tikz]{mdframed}
\usetikzlibrary{shadows}

\newmdenv[shadow=true,shadowcolor=black,rightmargin=8pt, skipabove=1em, skipbelow=1em]{shadedbox}

\usepackage{hyperref}
\usepackage[
type={CC},
modifier={by-nc},
version={4.0},
]{doclicense}

\usepackage[absolute,showboxes]{textpos}
\TPMargin{5pt}
\textblockcolour{gray!02}

\newcommand{\copyrightstatement}{
	\begin{textblock}{0.8}(0.095,0.02)    
		\noindent
		\scriptsize
		\copyright \  
		2021 held by the authors. This preprint is an accepted version to be published as an OpenAccess article in an upcoming issue of ACM Computing Surveys (CSUR). The final, published version of this article will be available under DOI \url{http://dx.doi.org/10.1145/3462513}.  \doclicenseThis
	\end{textblock}
}

\setcopyright{rightsretained}
\acmJournal{CSUR}
\acmYear{2021} \acmVolume{1} \acmNumber{1} \acmArticle{1} \acmMonth{1} \acmPrice{}
\acmDOI{10.1145/3462513}

\begin{document}
\copyrightstatement
\title{A Survey on Resilience in the IoT: Taxonomy, Classification and Discussion of Resilience Mechanisms}

\author{Christian Berger}
\authornote{Responding author}
\email{cb@sec.uni-passau.de}
\orcid{0000-0003-2754-9530}
\affiliation{%
  \institution{University of Passau}
  \streetaddress{Innstraße 41}
  \city{Passau}
  \state{Germany}
  \postcode{94032}
}

\author{Philipp Eichhammer}
\email{pe@sec.uni-passau.de}
\orcid{0000-0003-0604-4437}
\affiliation{%
	\institution{University of Passau}
	\streetaddress{Innstraße 41}
	\city{Passau}
	\state{Germany}
	\postcode{94032}
}

\author{Hans P. Reiser}
\email{hr@sec.uni-passau.de}
\orcid{0000-0002-2815-5747}
\affiliation{%
	\institution{University of Passau}
	\streetaddress{Innstraße 41}
	\city{Passau}
	\state{Germany}
	\postcode{89081}
}

\author{J\"org Domaschka}
\email{joerg.domaschka@uni-ulm.de}
\orcid{0000-0002-5451-3480}
\affiliation{%
	\institution{Ulm University}
	\streetaddress{Albert-Einstein-Allee 43}
	\city{Ulm}
	\state{Germany}
	\postcode{89081}
}

\author{Franz J. Hauck}
\email{franz.hauck@uni-ulm.de}
\orcid{0000-0002-7480-9617}
\affiliation{%
	\institution{Ulm University}
	\streetaddress{Albert-Einstein-Allee 11}
	\city{Ulm}
	\state{Germany}
	\postcode{89081}
}

\author{Gerhard Habiger}
\email{gerhard.habiger@uni-ulm.de}
\orcid{0000-0001-5603-3838}
\affiliation{%
	\institution{Ulm University}
	\streetaddress{Albert-Einstein-Allee 11}
	\city{Ulm}
	\state{Germany}
	\postcode{89081}
}

\renewcommand{\shortauthors}{C. Berger, P. Eichhammer, H. P. Reiser, J. Domaschka, F. J. Hauck, G. Habiger}

\begin{abstract}
	Internet-of-Things (IoT) ecosystems tend to grow both in scale and complexity as they consist of a variety of heterogeneous devices, which span over multiple architectural IoT layers (e.g., cloud, edge, sensors). 
Further, IoT  systems increasingly demand the resilient operability of services as they become part of critical infrastructures. This leads to a broad variety of research works that aim to increase the resilience of these systems.
In this paper, we create a systematization of knowledge about existing scientific efforts of making IoT systems resilient. In particular, 
we first discuss the taxonomy and classification of resilience and resilience mechanisms and subsequently survey state-of-the-art resilience mechanisms that have been proposed by research work and are applicable to IoT. 
As part of the survey, we also discuss questions that focus on the practical aspects of resilience, e.g., which constraints resilience mechanisms impose on developers when designing resilient systems by incorporating a specific mechanism into IoT systems.
\end{abstract}



\maketitle

%

\input{sections/intro.tex}

\input{sections/methodology}

\input{sections/taxonomy.tex}

\input{sections/measurability.tex}

\input{sections/mechanisms.tex}

\input{sections/related-work.tex}

\input{sections/conclusion.tex}

\begin{acks}
The authors thank the anonymous referees for their valuable comments and helpful suggestions.
This work has received financial support by the \grantsponsor{bmbf}{Federal Ministry of 
Education and Research of Germany}{} under grant no \grantnum{bmbf}{01IS18068, SORRIR} and
the \grantsponsor{dfg}{Deutsche Forschungsgemeinschaft (DFG, German
Research Foundation)}{} under grant no \grantnum{dfg}{268730775, OptSCORE}.
\end{acks}

\bibliographystyle{abbrv}
\bibliography{bibliography}

\end{document}

%% file: sections/intro.tex
\section{Introduction}

\begin{figure}[bt]
	\centering
	\includegraphics[width=0.635\columnwidth]{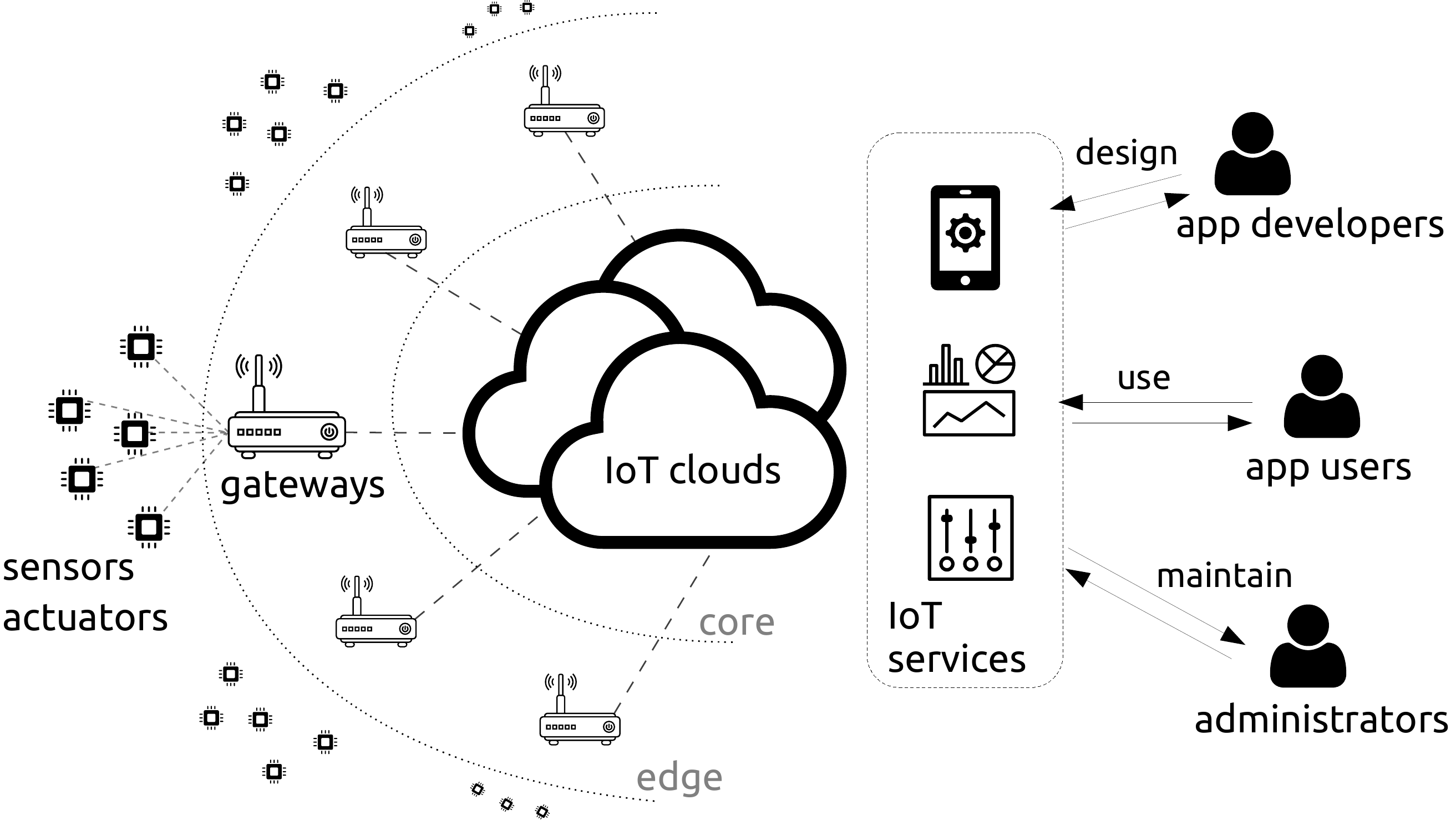}
	\caption{Layered IoT architecture~\cite{eichhammer2019towards}. 
	}
	\label{fig:iot_architecture}
\end{figure}

From a technical view, Internet-of-Things~(IoT) systems can be clearly distinguished from other systems by a set of technical characteristics. First, they
 tend to display a layered architecture where system components spread out across different layers, e.g., \textit{sensor landscape}, \textit{edge}, and \textit{cloud}~\cite{Taivalsaari19}~(see Figure~\ref{fig:iot_architecture}). Furthermore, IoT systems consist of a broad variety of heterogeneous system components. Typically, IoT components can be somehow classified, e.g., by computational power, capabilities, or locality to fit within one of the previously mentioned architectural layers.
 
IoT systems extend the world of computing to the domain of physical objects, which now become part of the overall system, e.g., by being equipped with sensors and actuators. Ideally, system operators want to ensure the longevity of their IoT systems and thus want their systems to preserve security and dependability properties by implementing  concrete resilience mechanisms. These mechanisms allow a system to display
behavioral stability (e.g., absorption, adaption, or recovery actions) when facing changes (such as disruptions like attacks or accidental faults). In particular, 
 a fault occurring in some individual component within one of the IoT layers should not lead to the system becoming unable to match its behavioral requirements.
System operators seek solutions and are assisted by a variety of  efforts that have been made to support the resilience of IoT systems. 

  
The goal of this paper is to first investigate whether there is a general view and a common understanding of resilience across academic literature, including the definition of resilience, its properties, and classifications. Subsequently, we shift the focus towards resilience in the IoT domain and outline which particular challenges exist here. We also discuss the practical aspects of resilience and review a set of resilience mechanisms for their applicability in IoT.

\subsection{Motivation}

A recent research roadmap has shown the need for research on resilience in IoT infrastructures~\cite{tsigkanos2019towards}.
The overall goal is often to secure and harden IoT systems against both hardware and software failures~\cite{terry16towards, Abreu2017resilient, Gia2015fault, Zhou2015supporting, Hasan2017optimizing, Christidis16blockchains} including sensor nodes, computers, communication links, and cloud services, and to protect against malicious attacks targeted at the system. Other works also address the challenge of scaling resilience with a growing number of possibly heterogeneous devices~\cite{resilienceIoTRoadmap2019}, and of matching requirements specified by system operators. 

In this document, we first want to examine if there is a common understanding of the \emph{definition of resilience}, and, in particular, which \textit{properties} of an IoT system are encompassed by the term resilience. 
Subsequently, we study \emph{means} for assuring these characteristics in more detail. 
We are using the term \textit{resilience mechanisms} for these means.

With resilience fitting the field of dependable and secure system design, the seminal work of Avi{\v{z}}ienis et al. marks a good starting point~\cite{avizienis04basic}. It gives precise definitions characterizing concepts of dependability and security of computing 
and communication systems. Building on their work, this survey shifts the focus to challenges of IoT systems and hence, generates a broader view on resilience covering topics that have not been sufficiently addressed by Avi{\v{z}}ienis et al.;
 particularly:

(1)~We categorize and elaborate on \emph{security mechanisms} beyond the set of intended security goals defined by Avi{\v{z}}ienis et al. who also disregard concerns such as privacy and identity management~\cite{avizienis04basic}. For IoT systems, which extend to the 
physical world and where components may collect or process personal data, privacy is a highly important matter: Consequently, neither software nor hardware components should leak information uncontrolled towards the outside world; if they do they cannot be considered to 
work as intended. Identity management and authorization services also play an important role in the IoT, since nowadays systems are more open and of a larger scale than systems were over a decade ago. Further, current systems tend to include more 
heterogeneity and are more dynamic~(rapidly evolving), making it even harder to keep track of all the components involved.

 (2)~We explicitly address IoT architectures which typically display a \emph{multi-layer architecture} divided into core, edge, and device landscape. IoT systems face various limitations and face heterogeneous requirements both of which complicate the implementation of resilience. We also provide more technical details on mechanisms discussing them from a practical and implementation-level view that is augmented with different examples of IoT systems.
 
 (3)~We extend their \emph{binary stance} on attributes to a continuum reflecting more conditions of the real world. For instance, in contrast to assuming that fault tolerance encapsulates mechanisms that can fully preserve the system functionality by compensating or removing a fault, we acknowledge that in practical settings such system should continue to function, probably in a degraded form, even when faults can no longer be compensated. IoT systems tend to explore new approaches for resilience by leveraging mechanisms that may work towards providing a ``best-effort'' resilience. 

\subsection{Research Questions}
This paper addresses research questions about the taxonomy and classification of resilience in general, and in particular by identifying resilience mechanisms that can be leveraged as building blocks to design resilient IoT systems. Overall, the questions we want to answer are the following:
\begin{enumerate}
    \item[\circled{1}] \textbf{Definitions of resilience and resilience mechanism:} 
     Does a consistent understanding of resilience in distributed systems, and in particular in IoT, exist? What is a common understanding of resilience, e.g., which concepts and attributes does resilience encompass? Which mechanisms can be labeled as resilience mechanisms?
    \item[\circled{2}] \textbf{Measurability, adjustability, and best-effort resilience:} Which approaches exist for quantifying resilience? What are relevant dimensions that need to be considered for quantification? How can system operators align systems to satisfy resilience requirements?
    Can we define degrees of preservation of system functionality? 
    Rather than only distinguishing between correct service execution 
     and service failure, how can a system be designed to deliver its service as a best effort, e.g., if faults  cannot be fully tolerated~(\textit{graceful degradation})?    
     	\item[\circled{3}] \textbf{Classification of resilience mechanisms:} Which aspects can be used for a taxonomy of resilience mechanisms in the IoT? More specifically: 
     \begin{itemize}
     	\item Depending on technical properties of IoT infrastructures, in which different layers~(e.g., sensor landscape, edge, cloud) do resilience mechanisms operate in?
     	\item What requirements must applications fulfill to support resilience mechanisms? 
     	Which considerations need developers  take into account when designing resilient IoT applications?
     \end{itemize}

\end{enumerate}

\subsection{Outline}
The remainder of this paper is outlined as follows: 
Section~\ref{section:methodology} summarizes the methodology employed when creating this survey. Subsequently, 
Section~\ref{sect:taxonomy} provides a more general taxonomy on resilience, including its definitions, properties, and classifications. Then, we shift our focus more towards resilience in the IoT. Further, we discuss the measurability and adjustability of resilience in Section~\ref{sect:measurabilityAndAdjustability}.
In Section~\ref{sect:resilience-mechanisms}, we analyze and discuss a broad range of resilience mechanisms from a more practical, implementation-level view, i.e., in respect to the constraints they may impose on applications.
Section~\ref{sect:related-work} gives an overview of related work. In Section~\ref{sect:conclusion}, we draw our conclusions.

%% file: sections/methodology.tex
\section{Methodology}
\label{section:methodology}
To create this survey, we employed 
 a basis of research papers that we initially knew and which covered a traditional view on resilience, by focusing on achieving security and dependability in distributed systems. A strong inspiration has been the survey of Avi{\v{z}}ienis et al.~\cite{avizienis04basic}. Building on this, we performed a systematic search that focused on resilience in the IoT.
To increase transparency and reproducibility, we briefly present our search methodology to gather and select a large body of academic literature in the field of resilience for IoT systems.
 Our main ambitions are the following:
\begin{itemize}
	\item Find distinct approaches in literature for defining the term \emph{resilience} to reason about the question whether a consistent understanding of this term exists
	\item Gather different approaches for measuring resilience aspects (which might be either quantitative or by ascertaining a specific quality of service can be provided)
	\item Identify and classify resilience mechanisms to create a broad taxonomy for resilience mechanisms in the IoT
\end{itemize}

 \textbf{Search strategy.} 
For literature research we used the ACM Digital Library with currently 609,508 records. Further, we employed the following search queries without filters on publication date. We combined the results of these queries and sanitized them for duplicates.

\begin{quote}
[[Publication Title: (iot)] OR  [Publication Title: "internet of things"]] AND [[Full Text: resilience] OR [Full Text: resilient]] (339 Results)
\end{quote}
 \begin{quote}
 [[Full Text: (iot)] OR [Full Text: "internet of things"]] AND [[Publication Title: resilience] OR [Publication Title: resilient]]  (88 Results)
 \end{quote}

 \textbf{Selection criteria.} 
 We used a list of selection criteria that can be applied to determine whether a paper found by the search queries should be selected to be included in the survey. These criteria also encompass conditions used to assess the quality of a found paper. 
 
 \emph{Inclusion criteria} -- A paper is included if one of the following criteria is satisfied:
 
 \begin{itemize}
\item The paper presents an own approach for defining, describing, or classifying the term resilience;
\item it elaborates metrics to reason about or measure resilience,
\item or presents a specific resilience mechanism that can be applied in at least one IoT layer.
 \end{itemize}

 \emph{Exclusion criteria} -- The exclusion process is applied after inclusion.
 A paper is excluded if only one of the following criteria applies:
 
  \begin{itemize}
 	\item The paper is not peer-reviewed; 
 	\item the paper presents its own view on resilience but belongs to another field (other than distributed systems);
 	\item the presented paper proposes a resilience mechanism but does not present a careful experimental evaluation showing how it can improve resilience;
 	\item  the paper lacks relevance: the paper presents a resilience mechanism but the presented mechanism is an incremental refinement of an earlier proposed mechanism.
 \end{itemize}

 \textbf{Selection procedure and results.} Our selection procedure is the following: The first phase consists of a fast scan, where each paper is examined by a single assessor to check if the paper could be of interest. In this phase, only papers that obviously do not qualify are sorted out, e.g., if the field is not computer science, the topic of the paper is not related to resilience, or the paper is not a research paper but a short abstract or workshop invitation. In the second round, all remaining papers are examined by two different reviewers to check for their relevance: A paper is relevant if at least one inclusion criterion applies and none of the exclusion criteria applies. Subsequently, all papers are being tagged either \emph{relevant} or \emph{not relevant}. In a final phase, conflicts among assessors are  resolved by including a third assessor and discussing the disagreement.
In the end, 41 papers have been selected for inclusion in our survey.

%% file: sections/taxonomy.tex
\section{Taxonomy}
\label{sect:taxonomy}

In order to clarify the taxonomy, we first give a broad overview of the usage and meaning of the term \emph{resilience} across different domains. 
We investigate whether there is a consistent understanding of this term in general and in the context of IoT. 
From these findings, we derive a definition for \emph{resilience} as well as for \emph{resilience mechanism} that fits
the IoT domain~(Section~\ref{sect:definitions}). 
Further, we investigate the different attributes and concepts resilience encompasses in general, i.e., in the broader 
field of distributed systems, to establish an understanding of this subject~(Section~\ref{sect:resilience-attributes}). 
Subsequently, we concretize and align our taxonomy of resilience mechanisms with the field of IoT by considering IoT-specific technical system properties and challenges~(Section~\ref{sect:classification-iot}).

\subsection{Definitions} 
\label{sect:definitions}
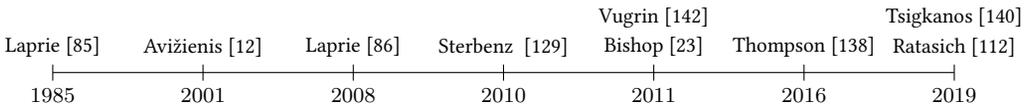
\begin{figure}[tb]
	\centering
	\begin{tikzpicture}[snake=zigzag, line before snake = 5mm, line after snake = 5mm]
	\draw (0,0) -- (12,0);
	
	\foreach \x in {0,2,4,6,8,10,12}
	\draw (\x cm,3pt) -- (\x cm,-3pt);
	\footnotesize
	\draw (0,0) node[below=3pt] {$ 1985 $} 
	node[above=3pt] { Laprie~\cite{laprie1985dependable} };
	\draw (2,0) node[below=3pt] {$ 2001 $} 
	node[above=3pt] {Avi{\v{z}}ienis~\cite{avizienis2001fundamental}};
	
	\draw (4,0) node[below=3pt] {$ 2008 $} 
	node[above=3pt] { Laprie~\cite{laprie2008dependability} };
	
	\draw (6,0) node[below=3pt] {$ 2010 $} 

		node[above=3pt] {Sterbenz
			~\cite{sterbenz2010resilience}};
	
		\draw (8,0) node[below=3pt] {$ 2011 $} 
		node[above=3pt] {Bishop~\cite{bishop2011resilience}}
		node[above=14pt] {Vugrin~\cite{vugrin2011resilience}};
		
	\draw (10,0) node[below=3pt] {$ 2016 $} 
	node[above=3pt] { Thompson~\cite{thompson16taxonomy}};
	
	\draw (12,0) node[below=3pt] {$ 2019 $} 
	node[above=3pt] { Ratasich~\cite{resilienceIoTRoadmap2019}} 
	node[above=14pt] {Tsigkanos~\cite{tsigkanos2019towards}};

	\end{tikzpicture}
	\caption{Timeline for definitions related to resilience.}
	\label{fig:resilience_definitions}
\end{figure}

In academic literature there is a multitude of definitions for the term resilience, mainly depending on the research 
domain. In addition to that, the attributes of resilient systems are often also lacking a consistent notion in 
literature. Hence, in this chapter, we try to find a common understanding of the term \emph{resilience}. Figure~\ref{fig:resilience_definitions} shows a brief timeline of selected definitions.

\textbf{A general view on resilience.} Thompson et al. collected  general definitions of resilience from various research fields, e.g., in the context of nature and environment, the public sector, and others~\cite{thompson16taxonomy}. One of it is from the US government giving an organizational view on resilience as

\begin{quote}
	\textit{the ability to adapt to changing conditions and prepare for, withstand, and rapidly recover from disruption}~\cite{dhs2010}.
\end{quote}

\noindent This definition is covering a general view on resilience and outlines that resilient systems need to prepare 
for~(possibly unpredictable) changes, and then be capable of re-entering a functional state again~(thus ``bouncing 
back''). This general notion of resilience describes the capacity of a system to maintain behavioral stability, which is 
also being elaborated by different concepts in the work of Vugrin et al. who describe resilience by three general 
concepts~\cite{vugrin2011resilience}: \textit{absorption}~(withstand disruptions), \textit{adaption}~(self-organize to, 
e.g., regain performance), and \textit{recovery}~(ability of a system to be repaired easily).
Bishop et al. understand resilience as the property that a system may degrade in its \textit{quality of service} under stress over time, but is able to recover~\cite{bishop2011resilience}:

\begin{quote}
\textit{It is important to note that such capability implies that the system must not cease to exist—that is, it
must survive at some capacity, in order to autonomously recover. In the cyber domain, a resilient system continues to
provide essential functionality, even under duress or in an
impaired state}~\cite{bishop2011resilience}.
\end{quote}
From a 
technical perspective, Thompson et al. provide a definition of security-related resilience:

\begin{quote}
	\textit{resilience is the maintenance of the nominated state of security}~\cite{thompson16taxonomy}.
\end{quote}

\noindent Here, the \textit{nominated state} is defined as a specific condition determined through a governance process that assesses the intrinsic value of the resource that is designated as requiring security~\cite{thompson16taxonomy}. 
Security is violated when the nominated state changed. This concludes to resilience being the ability to prevent or resolve that change, thus maintaining the nominated state and withstanding disruptions. 
In the field of computer networks, Sterbenz et al. define resilience as
\begin{quote}
	\textit{the ability of the network to provide
		and maintain an acceptable level of service in the face of various faults and challenges to normal operation}~\cite{sterbenz2010resilience}.
\end{quote}
Further, in the field of dependable systems, among oft-cited definitions are the ones given by Laprie et al. in their work \textit{From dependability to resilience}~\cite{laprie2008dependability}. They define \textit{resilience} as

\begin{quote}
	\textit{the persistence of service delivery that can justifiably be trusted, when facing changes}~\cite{laprie2008dependability}.
\end{quote}

\noindent This definition extends the definition of \textit{dependability} that has been used by Laprie in earlier works:

\begin{quote}
\textit{dependability is the quality of the delivered service such that reliance can justifiably be placed on this service}~\cite{laprie1985dependable}.
\end{quote}
Avi{\v{z}}ienis et al. also follow this view and define dependability as 

\begin{quote}
	\textit{the ability to deliver service that can justifiably be trusted}~\cite{avizienis2001fundamental}.
\end{quote}

\noindent Note that the definition of resilience from Laprie et al. reflects the ambition to preserve \textit{dependability} in the face of (possibly unforeseen) \textit{changes}. In particular, resilience means maintaining dependability even if unexpected things happen.
Laprie et al. also conclude this by providing a shorthand, alternate definition:

\begin{quote}
	\textit{the persistence of dependability when facing change}~\cite{laprie2008dependability}.
\end{quote}

\textbf{Changes and disruptions.} So far, these definitions employ the terminology of \textit{disruptions} and \textit{changes}. A disruption is a very specific form of change. 
According to Tsigkanos et al., \textit{disruption is an adverse change to system stability} and it can fundamentally
affect system requirements~\cite{tsigkanos2019towards}. Adverse changes can be either \textit{external} to the system, e.g., caused by the environment the system operates in, or \textit{internal}, such as when some internal fault occurs. Such changes might push the system into an unforeseen and possibly unwanted state. On the other hand, changes can be intended and desired as well, e.g., to improve or evolve the system. Planned changes are part of a systematic and subsequent development of the system, such as adding additional resources or novel application components to a system.

 
\textbf{Resilience in IoT.} 
One concrete goal of our literature research was to find out which 
approaches for defining resilience in the IoT domain exist and whether the understanding of this term is consistent.
After applying our search methodology and investigating which understanding IoT papers have under the term resilience, we made a few observations. First, a large body of papers uses the adjective \textit{resilient} rather nonchalantly to characterize their systems or solutions without explicitly providing an own definition or referencing an existing definition. The attribute \textit{resilient} is just added without further explanation and its interpretation is left to the reader. 
Second, in other papers, \textit{resilient} is used in \textit{conjunction}, e.g., a developed approach is resilient \textit{against}  specific attacks, faults, perturbations, or overloads. In this type of papers (e.g., \cite{yong2018switching, jin2019resilient, beal2017self, benson2018firedex}), the understanding of resilience is implicitly made clear as withstanding specific system alterations (e.g., attacks, faults, or overload) to maintain a functional system state. Third, a few IoT papers explicitly define their own resilience definition or employ existing definitions. 
We discuss these different approaches to defining resilience in the following.

\noindent Delic et al. define resilience by referring to a system's \textit{behavioral} requirements of being capable of self-stabilizing its state during and after change:

\begin{quote}
\textit{The	 resiliency of a system	is defined	by its	capability (1) to resist external perturbances and	
internal failures; (2) to recover and enter stable state(s); and (3) to adapt its	 structure and	
behavior to constant change}~\cite{delic16onResilienceofIoT}.
\end{quote}

\noindent In the context of smart cities, Modarresi et al. adopt the definition of Sterbenz et al.~\cite{sterbenz2010resilience} and give a \textit{goal-oriented} view on resilience, that is, maintaining an acceptable level of service during changes:

\begin{quote}
\textit{We define resilience as the ability of the system to provide and maintain an acceptable level of service in the face of various faults and challenges to normal operation}~\cite{modarresi17multilevel}.
\end{quote}

\noindent Pradhan et al. present a goal-driven orchestration middleware for resilient IoT systems. They employ their own explanation of resilience. In their view resilience is goal-oriented and has behavioral requirements:

\begin{quote}
\textit{Each application deployed for a mission has specific goal(s) that must be
satisfied at all times. IoT systems should therefore be equipped with mechanisms that ensure all critical goals are satisfied for as long as possible, i.e., they must be resilient by facilitating failure avoidance, failure management, and operations management to support incremental hardware and software changes over time}~\cite{Pradhan18chariot}.
\end{quote}

\noindent Further, Khan et al. use a resilience definition for routing in the IoT~\cite{khan17trust} that is borrowed from earlier work~\cite{Erdene-Ochir12new}. This definition reasons about both the behavior~(absorption capability) and goal~(continuation of delivery) of resilience: 

\begin{quote}
\textit{the ability of a network
\textit{to absorb the performance degradation under some failure pattern
(random or intentional) and to continue delivering messages with an
increasing number of $k$ compromised nodes}}~\cite{Erdene-Ochir12new}.
\end{quote}

\noindent Another definition focusing on service continuation despite changes is given by Witti et al.:

\begin{quote}
\textit{Device resilience refers [to] the ability of a component to maintain
service with alteration in the system environment}~\cite{Witti18secure}.
\end{quote}

\noindent Since IoT systems are systems that tend to evolve in larger scale and have considerable dynamics, complexity, and heterogeneity of involved components, these circumstances may be regarded in a resilience definition as well: 
Ratasich et al. extend Laprie's definition and argue that resilience is a property which, in the context of IoT systems, should also \textit{scale} dependability and security, e.g., when dealing with environmental or technological changes, which they refer to as \textit{long-term dependability and security}~\cite{resilienceIoTRoadmap2019}.
Here, the definition of dependability is taken from Avi{\v{z}}ienis et al. who define dependability as \textit{the ability to deliver service that can justifiably be trusted}~\cite{avizienis2001fundamental}. 
 Avi{\v{z}}ienis also follows the definition of Laprie et al. in his more recent work~\cite{avivzienis2017visit}. 
Tsigkanos et al. summarize a vision, challenges and research directions road-map for IoT systems, and they define resilience as 
	\textit{the
		persistence of reliable requirements satisfaction when facing
		change}~\cite{tsigkanos2019towards},
thus largely following the view of Laprie, too. 

\begin{figure}[t]
	\includegraphics[width=0.97\textwidth]{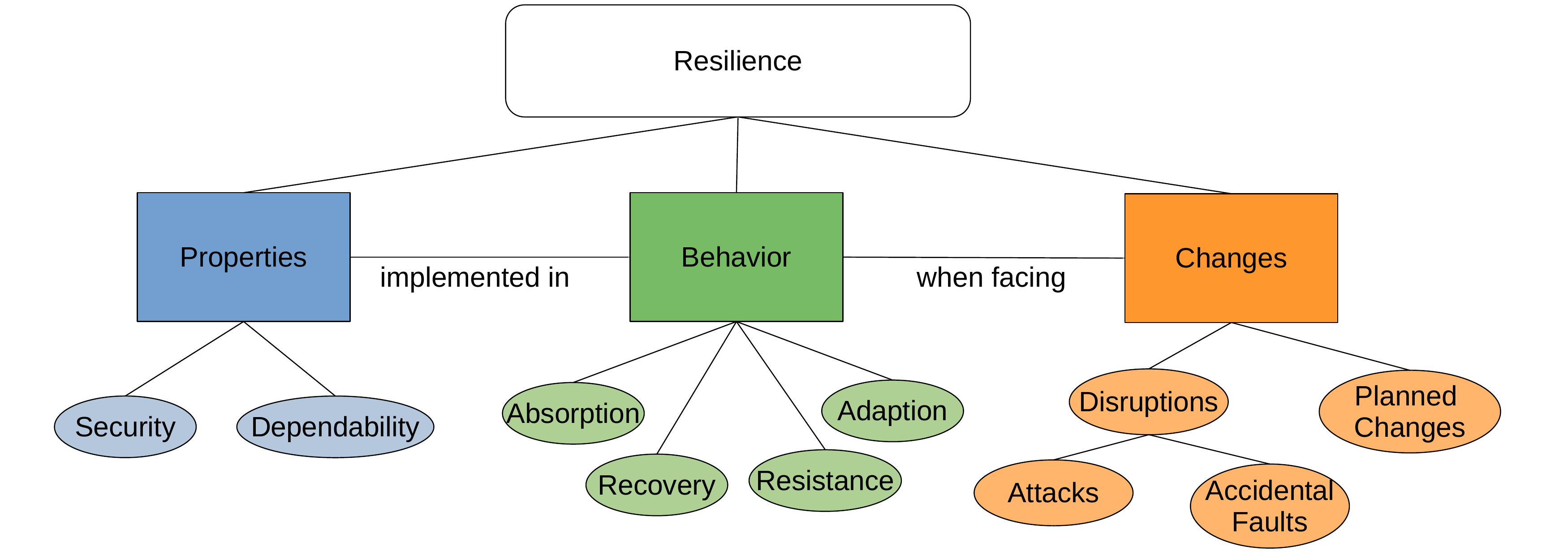}
	\caption{Resilience terms and their relation. Resilience can be understood as the ambition to preserve dependability and security properties when facing changes, which, in turn, manifests in a specific behavior of the system that facilitates stability despite these changes. }
	\label{fig:resilience_terms}
\end{figure}

\textbf{Summary and take-away message.} Summarized, we see that the definitions of resilience require the system to be
 capable of, \textit{in the face of changes}, preserving some correct state and functionality (``withstanding changes'') or 
 to provide means to recover and go back to an intended state (``bouncing back''). Being able to maintain behavioral 
 stability of a system even during disruptions is a core characteristic of resilience. We conclude that, while the 
 terminology used to define  resilience is diverse in literature, the understanding of resilience is substantially consistent: 
  If the goal of a resilient system is to be capable of delivering service that can justifiably be trusted even when 
  facing changes, then it needs to provide mechanisms for ensuring behavioral stability, that is, absorbing disruptions,
  temporarily degrading its state but subsequently recovering~(bouncing back) to its nominated state later on.
  Moreover, in the IoT domain, we found that the definitions of Laprie and Sterbenz seem to be rather popular as they have both been employed (either one-to-one  or in an adapted form) in several IoT papers~\cite{laprie1985dependable,sterbenz2010resilience}.
   In Figure~\ref{fig:resilience_terms}, we show a diagram that illustrates how terms used to define resilience are connected. 

For our own definition of resilience we strongly identify with the ones of Laprie et al., hence, also largely with Avi{\v{z}}ienis et al. and Ratasich et al.. Further, we also distill security as a main property to resilience besides 
dependability. In addition to outlining the properties dependability and security, we also \textit{emphasize the expected behavior} of resilient systems, that is, being able to withstand changes or recover. 
In summary, our definitions of resilience and resilience mechanisms are formulated as:

\begin{shadedbox} 
	\textit{Resilience}
	 is the property of preserving the dependability and security of a system when the system encounters changes, thus withstanding or recovering from impairments. 
	  \textit{Resilience mechanisms} are all means that work towards achieving this property.
\end{shadedbox}

We emphasize our understanding of \textit{resilience} as an umbrella term, which considers both dependability and security properties, thus asserting that some well-defined correct state can be recovered or preserved (possibly temporarily at some degraded level) even if changes, such as (potentially malicious) faults and attacks, occur in the system.
Consequently, mechanisms that ensure the persistence of dependability and security when facing changes are \textit{resilience mechanisms}.

\textbf{Fault, error, and failure.} 
When we speak about faults in the remainder of this paper, we stick to the terminology introduced by Avi{\v{z}}ienis et 
al., in particular, we employ the \textit{fault}$\rightarrow$\textit{error}$\rightarrow$\textit{failure} 
chain~\cite{avizienis04basic}: \textit{Faults} can be dormant or active. Once a fault activates, it produces an 
\textit{error}, hence a fault is the cause of some error,  while an error is the manifestation of some preceding fault, e.g., in form of an incorrect state. If an error leads to another error, we call this \textit{error propagation}. 
Further, if some error
leads to a deviation of system behavior from its intended service delivery, e.g., a system produces and returns a wrong result because of its own erroneous state, then we call it a \textit{failure}.  
Note that this chain may propagate recursively, because the failure of some component $A$ may appear as an external fault to another component $B$, if the correct execution of $B$ depends on receiving correct results from $A$. For a more specific classification of fault types, we refer to the extensive taxonomy of Avi{\v{z}}ienis et al.~\cite{avizienis04basic}.

  \textbf{Relation between resilience and security.}
Often, research works covering the topic of resilience mainly focus on dependability mechanisms, e.g., fault tolerance and fault prevention.
Sometimes, resilience and security are even treated as two separate, partly distinct subjects. 
From the view of a system operator, however, shifting the focus between dependability and security might depend on
knowledge about the environment a system runs in and the capability of making reasonable assumptions about the behavior and interactions of the system with its users.
If available, these assumptions can be used for creating models~(e.g., a fault model or an attacker model) and to 
subsequently select appropriate mechanisms that work for the selected model. Resilience consists of dependability and 
security. Both are often closely related as they share a common set of system properties, such as ensuring availability and integrity~\cite{avizienis04basic}. 
Thus, some mechanisms can also fit into both categories. An example of this is intrusion-tolerant 
replication~\cite{bessani2011byzantine}, where the integrity and availability of a service is preserved even if a 
bounded amount $f$ of replicas becomes faulty or intruded\footnote{Generally, attacks like an intrusion could be viewed 
as malicious, external faults.} by an attacker within a bounded amount of time $T$, assuming the system employs proactive recovery~\cite{sousa2009highly}. 
Our definition of resilience emphasizes that resilience also needs to preserve the security of a system in the face of 
harmful conditions like attacks, 
thus resilience includes security.

\subsection{Attributes and Basic Classifications}
\label{sect:resilience-attributes}
Section~\ref{sect:definitions} shows that there is a broad variety of definitions for resilience. In consequence, the properties and concepts used to describe a resilient system are varying as well and moreover sometimes become overloaded with different meanings.
In this section, we seek to find a common set of well-defined attributes, which are both applicable for practical IoT systems and also commonly agreed upon across different research works. 
Due to the fact that definitions of resilience are often linked to the properties \textit{dependability} and \textit{security}, we begin by considering resilience attributes as the set of attributes that are the conjunction of the security and dependability attributes. As a starting point, we refer to the well-established taxonomy of Avi{\v{z}}ienis et al., who characterize \textit{dependability} as consisting of the following attributes~\cite{avizienis04basic}:
\begin{itemize}
	\item \textbf{availability} \textit{(readiness for correct service)}
	\item \textbf{integrity} \textit{(absence of improper system alterations)}
	\item \textbf{reliability} \textit{(continuity of correct service)}
	\item \textbf{safety} \textit{(absence of catastrophic consequences)}
	\item \textbf{maintainability} \textit{(ability to undergo modifications and repairs)}
\end{itemize}
 Furthermore, \textit{security} includes \textbf{availability}, \textbf{integrity}, and 
 \begin{itemize}
	\item \textbf{confidentiality} \textit{(absence of unauthorized system disclosure of information)} 
 \end{itemize}

\noindent The relation of system attributes as described by Avi{\v{z}}ienis et al. is largely agreed upon and has been adopted by a broad variety of academic literature in the field of resilient systems~\cite{resilienceIoTRoadmap2019, shirazi2017extended, cho2016metrics, bishop2011resilience, vieira2012resilience, sterbenz2010resilience}.

Yet, in this paper, differing from Avi{\v{z}}ienis, we consider \textit{availability} to be an attribute encompassing both maintainability and reliability.
 For instance, the probability for a system or component to keep working successfully up to a specified amount of time 
 can describe the property \textit{reliability} among other possible metrics. Generally, reliability can be modeled 
 through the use of stochastic means and metrics, e.g., mean time to failure.
Further, maintainability is a characteristic that expresses how easy (and time consuming) it is to repair a system in case of failures and thus to make it available again. Both attributes compose the availability of a system. Maintainability also considers planned changes to the system, e.g.,  by adding or removing components as the system evolves over time. Our approach of classifying availability matches the current IEC standard IEC 60050-192:2015 which notes that “\textit{[a]vailability depends upon the combined characteristics of the reliability, recoverability, and maintainability of the item, and the maintenance support performance}”~\cite{iec60050_192}. The minor difference here is only that we use “maintainability” as the generic umbrella term that encompasses the recoverability, the maintainability, and the maintenance support from the IEC definition.


\begin{figure}[t]
	\includegraphics[width=1\textwidth]{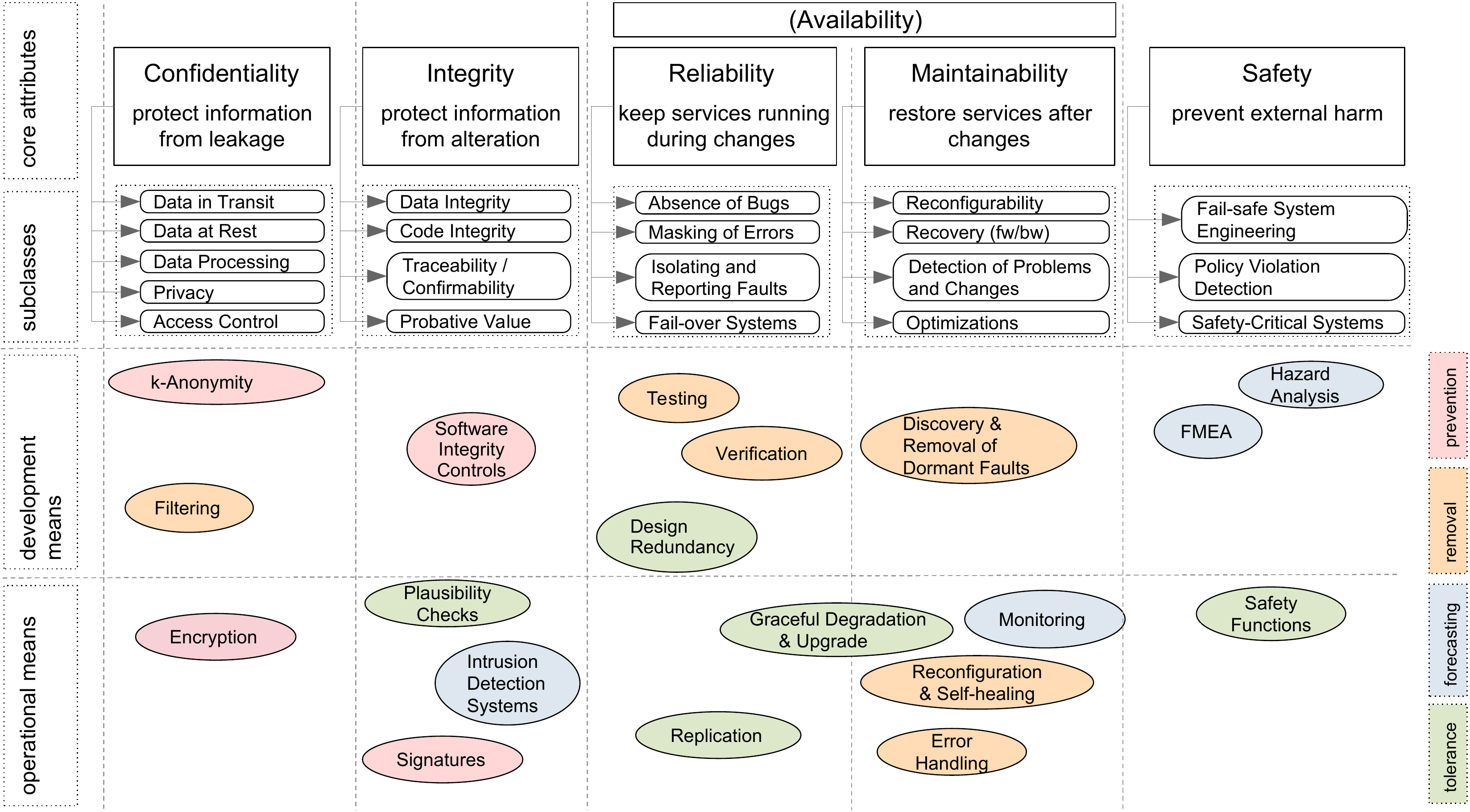}
	\caption{Attributes of resilience and means.}
	\label{fig:resilience_attributes}
\end{figure}

\textbf{Subclasses of resilience core attributes.} The core attributes can be further decomposed into different subclasses as shown in Figure~\ref{fig:resilience_attributes}.
The attribute \textit{confidentiality} consists of different subclasses that have the common goal of protecting information from leakage. This includes protecting data in transit, at rest, and while processing as well as controlling the legitimized access to data. When applications collect and process data related to individuals, e.g., a person's location, then privacy, which is the ability of an individual or group to express themselves selectively, also becomes a crucial concern. 

Further, \textit{integrity} is a property that predicates whether information or resources are protected from unauthorized modification. This includes the integrity of data, regardless of whether it is stored or in transit (e.g., in form of messages) as well as the integrity of applications (which process data), e.g., by ascertaining code integrity. In respect to users interacting with a system or with other users, it is often necessary to guarantee traceability, i.e., being able to trace specific actions back to users as well as the attached probative value, e.g., being able to prove the judicial liability of a specific user for some specific action.

\textit{Reliability} is a property that aims to make a promise about the system's correct behavior over time, such as ``the mean time for this component to fail is X days'' or ``the probability of a component failing before time T is reached is X''. The attribute \textit{reliability} encompasses, for instance, the absence of bugs, masking of errors, or isolating or reporting faults. The common goal among these is to keep the service running, despite of considering different paths of how this can be achieved. 

\textit{Maintainability} is an attribute which is often regarded in a broader sense. Overall, it means to be capable of restoring services after changes or problems occur. Firstly, if we consider \textit{adverse} changes, such as faults or attacks, we refer to the system's capability to react to these changes.
 In this respect, it encompasses the detection of such changes and problems and also the recovery of the system as well as its reconfiguration. Secondly, if we consider planned changes, such as adding new system components to provide additional functionality, or scaling the system size to increase its performance, maintainability may largely refer to how easy such an intended reconfiguration can be implemented by administrators.

Finally, \textit{safety} is a property that states that the system should prevent external harm to its environment or persons. As subclasses it encompasses fail-safe system engineering, policy violation detection, and safety-critical systems.

\textbf{Implementing resilience into practical systems.} There is a variety of means to implement resilience attributes into practical systems (some examples are shown in Figure~\ref{fig:resilience_attributes}). These means can be categorized into \textit{operational means}, which function during the operation of the system, and \textit{developmental means}, which are applied during the design phase of the system~\cite{avizienis04basic}. Moreover, means can be also classified further in respect to how they approach faults, into \textit{removal}, \textit{prevention}, \textit{tolerance}, and \textit{forecasting} types~\cite{avizienis04basic}.
Fault-removal means are employed to reduce faults, e.g., \textit{software testing} can discover bugs, which are then subsequently fixed. 
Moreover, fault-prevention means try to prevent the occurrence of faults, e.g., \textit{anonymization} techniques like k-anonymity~\cite{sweeney2002k} can protect a service from privacy-related faults.
Fault-tolerance means prevent service failures despite occurring faults, for example by masking the presence of faults using a replication protocol.
A typical example of this are replication protocols.

Finally, fault-forecasting means have the ambition to estimate both the present and future occurrences of faults and their consequences~\cite{avizienis04basic}. 
We show for some chosen resilience means how they fit into these categorizations in  Figure~\ref{fig:resilience_attributes}. It is noteworthy that a single resilience means can \textit{preserve multiple resilience attributes}, which is not sufficiently covered by Figure~\ref{fig:resilience_attributes}. For instance, \textit{replication} is aimed to preserve the availability of a service, e.g., by managing redundant copies of an application state, but could also protect integrity, by letting replicas redundantly execute requests with a majority agreeing on what is the correct copy of the application state.

While the previously presented core attributes are very commonly agreed upon across literature, there are also works that refine or extend this basic set.
Ratasich et al.~\cite{resilienceIoTRoadmap2019} unify the concepts of dependability and security as defined by Avi{\v{z}}ienis under the term \textit{resilience}. They also define the property \textit{scalable resilience}~\cite{resilienceIoTRoadmap2019}, which encompasses both \textit{resilience} (with its usual attributes) and \textit{evolvability}, which expresses the need of a heterogeneous and complex system to
be capable of evolving over time. This requires
dependability and security attributes that are considered during design time to
scale up with actual IoT system requirements.
Sterbenz et al. divide resilience properties into two groups~\cite{sterbenz2010resilience}: \textit{trustworthiness resilience} encompasses common \textit{security} and \textit{dependability} attributes along with \textit{performability}, which contains quality of service metrics, while the other group called \textit{challenge tolerance disciplines} consists of survivability (e.g., in case of correlated failures or disasters), disruption tolerance (e.g., caused by the environment) and traffic tolerance (such as DDos attacks).

\begin{shadedbox} 
	\textit{Summary.} The \textit{resilience attributes} of IoT systems cover the preservation of \textit{integrity}, \textit{confidentiality}, \textit{availability}, \textit{reliability}, \textit{maintainability }and \textit{safety} of these systems. Resilience mechanisms target the implementation of these properties in practical systems during development and operation. IoT systems tend to face new and unforeseen challenges as they grow in size, heterogeneity, and complexity over time. 
\end{shadedbox}


\subsection{IoT-specific Challenges for Resilience Mechanisms}
\label{sect:classification-iot}
After defining the term resilience and identifying its corresponding core attributes, we here shift the focus towards resilience in IoT, investigate challenges that exist for the design of resilience mechanisms, and further touch on possible solutions. We go into more detail on these in Section~\ref{sect:resilience-mechanisms}.

 The main challenges we identified for incorporating resilience mechanisms into IoT systems are \textit{architectural} challenges (e.g., device constraints, heterogeneity, and scalability), the challenge of limiting \textit{dependencies across layers},
 and \textit{administrative} challenges~(e.g., the dispersion of system components into different administrative domains). 

\textbf{Architectural challenges of IoT systems.}
IoT systems have a broad heterogeneity of physical components distributed across different layers within IoT infrastructures, namely sensor landscape~(device level), edge, and cloud. 
The heterogeneity of components results in a distinct variation in terms of available hardware resources such as computing capabilities, power constraints, and available memory. 
These also impose limitations on the applicability of certain resilience mechanisms~(we will discuss this in more detail in Section~\ref{sect:resilience-mechanisms}).
Further,  IoT system developers are also faced with
the question of how the overall system can be designed resiliently when spanning different layers, in contrast to just making individual components within a single layer resilient. 
Initially, among sensor landscape, edge, and cloud, different resilience mechanisms might be employed individually to preserve overall resilience properties for the system. 
For example, resilience techniques that preserve \textit{integrity} in the sensor landscape might include a design that uses redundancy among sensors~\cite{terry16towards}~(and ensures correctness by voting on measurement values in the edge, e.g., taking the median of values as measured by distinct sensors), plausibility checks~\cite{kuemper2018valid}~(e.g., does the reported value match some pre-defined expression, pattern, or range), and physical protection of sensors so that they can not be tampered with by malicious attackers~\cite{andrea2015internet} (e.g., holding a lighter towards a temperature sensor makes it produce wrong results). 
More sophisticated integrity-preserving mechanisms could be deployed in edge and cloud, e.g., intrusion-detection systems are often rather placed on edge nodes such as gateways~(e.g., because this allows observing the network flow) or cloud servers~(where data is often stored and processed), than on resource-constrained devices. 
Yet, some lightweight IDS solutions exist that enable IDS functionality even on small IoT devices~\cite{oh2014malicious, lee2014lightweight}. In this context, a technique that proves to be especially useful in the cloud to build IDS is \textit{virtual machine introspection}~\cite{garfinkel2003virtual}. Summarized, we need to carefully regard the technical limitations dictated by the IoT architecture when selecting and incorporating resilience mechanisms into system components.
 
\textbf{Resilience mechanisms should work across IoT layers.}
We see that it is possible to place resilience mechanisms into different layers and subsequently achieve certain resilience properties within some layer; however, we also need to be concerned with the dependencies across layers, and hence also design techniques that work across layers.
For instance, IoT systems might face connectivity disruptions between edge and cloud. 
Ideally, a resilient system should try to continue delivering its service functionality on a best-effort basis. 
By employing \textit{partition-tolerant redundancy} we could allow data that is usually stored and processed on cloud nodes to be also maintained (i.e., cached and processed) locally by edge nodes~\cite{javed2018cefiot}. 
As long as the connection between edge and cloud works, they can synchronize and keep data consistent. 
If the connectivity breaks, the system could degrade gracefully and fall back to letting the edge service operate under restricted performance or functionality. 
Here, the consistency between edge and cloud could temporarily be sacrificed in favor of keeping the service available. 
For instance, parts of the application state can be buffered in the edge and updated to the cloud at a later time. 
Several current research works provide fault-tolerance and availability-enhancing solutions between edge and cloud~\cite{chang2014bringing, javed2018cefiot, xiong2018extend, javed2020iotef}.
Apart from replication and redundancy, there are other resilience mechanisms that function across several IoT layers, 
such as monitoring mechanisms. 
In general, the goal of monitoring is, in most cases, the observation and analysis of as much information as possible to gain as detailed insights into an infrastructure as possible. 
In terms of IoT infrastructures, this means that data collection is performed on
 all possible devices across all layers. The data aggregation and analysis process is then carried out, depending on the monitoring mechanism, on various layers. 
 This distribution of mechanisms automatically enables fallback to local execution in case of failures. As an example, 
 let us assume a smart city operates several parking garages all collecting and analyzing monitoring data. In the event 
 of one garage losing connection to the central monitoring instance, it can still perform its local monitoring tasks, 
 e.g., IDS within its infrastructure. Yet,  incidents cannot be reported to other garages. Several current research efforts are focusing on distributing monitoring mechanisms from the cloud to the edge and device layers~\cite{breitenbacher19,ibrahim16}.

  We conclude that resilient IoT systems, and in particular the employed resilience mechanisms, should be designed in a way that they are agnostic of operating across layers, minimize component dependencies to adjacent layers and account for unpredictable faults occurring in components located in other layers.
  
\textbf{IoT systems can have multiple, different administrative domains.} 
For IoT systems, it is also important to distinguish between the technical and the  administrative view of the system.
For instance, an IoT system can consist of several administrative domains with disjunct administrators being in charge of specific IoT components and devices.
According to Tsigkanos et al., one important challenge for IoT is \textit{addressing mobility and distribution of software components between diverse administrative domains and locales}~\cite{tsigkanos2019towards}.
As an illustrative scenario, consider a smart city where available parking space is managed by an application that steers car drivers to a nearby parking garage with available space.
This application might span several parking garages, and every garage runs its own local management services.
Apart from that, there is a navigation service that maintains the global perspective of available parking space across the city and offers navigation services to drivers.
This requires \textit{coordination}, e.g., with a central cloud-based service provided and maintained by the city administration. 
Overall, these garages might be managed by different operators and administrators and the hierarchically superordinate coordination service also lies in some distinct administrative domain.
Every administrative domain might have its own resilience requirements and the need to harden itself for failures of components located in other domains---a centralized coordination service might represent a single point of failure in the worst case. 
Tsigkanos et al. argue that in order to increase resilience in IoT, it is necessary for software components to coordinate in a decentralized way and be capable of self-adaption when disruptions occur at runtime~\cite{tsigkanos2019towards}.
Overall, multiple and diverse administrative domains induce different challenges that need to be addressed by technical solutions and also need to be considered when designing resilience mechanisms. 
For instance, in the context of \textit{maintainability}, again the question arises how available objects in IoT can be efficiently coordinated and managed across different administrative domains~\cite{kyriazis2013smart}. 
In particular, diverse trust assumptions and identities may have to be considered and managed. 
Further, \textit{privacy} concerns emerge, e.g., when data is forwarded to possibly unknown administrative domains where privacy-preserving policies are neither known nor enforced.

%% file: sections/measurability.tex
\section{Measurability and Adjustability of Resilience}
\label{sect:measurabilityAndAdjustability}
In this section, we first investigate the question of how resilience can be measured and quantified, i.e., what different types of \textit{resilience metrics} exist and how they can be applied to practical IoT systems in Section~\ref{sect:measurability}.
Closely related to the question of quantifying resilience, and arguing about qualitative properties of a system, is the question of how \textit{adjustability} can be accomplished from the perspective of a system operator who wants to align his system to match specific resilience requirements. 
We discuss the latter question in Section~\ref{sect:adjustability}.

\subsection{Resilience Metrics}
\label{sect:measurability}

In the following, we identify several \textit{different} approaches to \textit{specifying or measuring} resilience. The reasons for this are mainly, that (i)~a broad variety of resilience mechanisms exists which have different goals and limitations and (ii)~mechanism-specific metrics
are applied to individual resilience-enabling approaches or a family of resilience mechanisms to subsequently argue about, e.g., their cost or effectiveness. We argue that resilience itself can not be treated as a single aspect and measured using some specific function. Rather, resilience encompasses a broad variety of solutions, often domain specific, that pursue a common goal. Resilience can best be understood as a multi-dimensional property of a system and thus multiple different metrics can be applied for resilience mechanisms, too. The selection of these metrics also depends on the resilience properties, which a resilience mechanism preserves and which should be quantified.

\textbf{Different goals mean employing different metrics.} 
Overall, the goal is often to reason about the \textit{effect} that a resilience mechanism has once applied to some system while using an underlying model to capture the granularity of fault types or malicious misbehavior that a system can then endure, e.g., the proportion of faulty components. 
For instance, if we were to argue about the \textit{effectiveness} of a resilience mechanism that should preserve the reliability of a system, then we could employ the \textit{mean time to failure} as a measure.
However, system engineers are usually concerned with preserving several, if not all, resilience properties, which in turn, requires having a multi-dimensional view of resilience and thus employing a toolbox of different metrics. 
In the following we review general and established \textit{concepts} for measuring resilience.

\textbf{Coverage.} Avi{\v{z}}ienis et al. define the \textit{measure of effectiveness} of some specific fault-tolerance technique as the \textit{coverage} or \textit{fault-tolerance coverage}~\cite{avizienis04basic}. 
This concept encompasses (i)~the \textit{error- and fault-handling coverage}, which is a measure to capture how many of the actually occurring faults are detected and treated by the fault-tolerance mechanism since its capabilities to handle actual faults might be restricted, e.g., due to development faults, and 
(ii)~the \textit{fault-assumption coverage} which is a measure for congruence between model and reality.
In particular faulty components can display behavior different from what is assumed by the fault-tolerance mechanisms (e.g., Byzantine instead of crash faults, correlated faults instead of independent ones).
Hence, these faults would not be successfully treated.
This measure indicates that the design of the fault-tolerance mechanism does not match reality fairly enough. 
An often made assumption in distributed systems and, in particular, replicated systems, is that individual components are assumed to fail independently. 
In practice, this needs to be regarded with caution as common bugs in these components can lead to correlated failures. N-version programming~\cite{chen1978n, avizienis1985n} could approach common software bugs  but is often viewed as impractical. However, reasonable models and reliability  metrics~\cite{dai2004model} to assess the overall system's reliability for these approaches exist.
For replicated systems, an interesting measure is the risk of having common weaknesses or vulnerabilities among replicas~\cite{garcia2019lazarus} and thus measure the configuration risk of a system which is useful to reason about the diversity of a state-machine replication (SMR) system~\cite{da2018diversity, garcia2019lazarus}

\textbf{Degree of replication and fail-over time.} A commonly used resilience mechanism that allows a system to tolerate crashing components is the \textit{primary-backup approach}~\cite{budhiraja1993primary}, providing a number of available and consistently updated backup replicas. In case the primary fails, a backup can replace the primary and continue service operation. 
Budhiraja et al. use the term~\textit{degree of replication}~\cite{budhiraja1993primary} mainly as a cost metric for 
counting redundancy, e.g., the concrete number of servers that are necessary to implement a resilience 
mechanism into a service. 
The authors also introduce a metric for limitations such as the \textit{fail-over time}, which is defined as \textit{worst case period during which requests can be lost because there is no primary}~\cite{budhiraja1993primary}.
This is a metric that upper-bounds the effort~(as measured in time) of the resilience mechanism to restore the 
functional state of the service and hence allows to give certain quality-of-service level guarantees to applications~(or developers building applications). The degree of replication is also present in active replication  protocols~\cite{schneider1990implementing}.
Replication protocols compensate the behavior or individual faulty components and usually specify and \textit{denote the number of faulty replicas} that can be tolerated as $f$ out of $n$ total replicas. It is noteworthy that the expenses of replication then also depend on the concrete fault model which is employed.
A fault model defines the assumptions with respect to which faults are considered in the system. For instance, in state-machine replication,  the \textit{Byzantine fault model}~\cite{lamport1982byzantine}, which assumes arbitrary behavior of faulty components, usually requires the BFT protocol to employ at least $n = 3f + 1 $ replicas, while it is only $2f+1$ in a crash fault-tolerant~(CFT) SMR protocol. If more than $f$ replicas fail at once it will bring the system to a state in which it can not continue to operate any longer. 

\textbf{General metrics.} 
Strigini et al. investigate general metrics and assessments of resilience in the field of engineering~\cite{strigini2012fault}. 
They list several metrics of \textit{tolerable disturbances}, which capture different ambitions of quantifying the resilience of a system, such as (i)~\textit{how far the object of interest can be pushed without losing its ability to rebound or recover}, e.g., how many replica failures could be masked in some replication protocol, (ii)~\textit{how quickly it will rebound}, thus measuring the time needed to re-obtain the ability to deliver a certain service, also referred to as \textit{failover time}~\cite{budhiraja1993primary} by Budhiraja, or (iii)~\textit{how closely its state after rebounding will resemble the state before the disturbance}, which is of particular interest if the resilience mechanism implements some sort of \textit{service degradation}.

\input{diagrams/radarchart.tex}

Note that even a high-level comparative statement on two systems in regard to resilience, such as ``system $A$ is more resilient than system $B$'' is very problematic because of the multi-dimensionality of objectives of resilience.
This calls for the application of different resilience metrics at once.

\textbf{Example.} Imagine we want to ensure the availability of some specific service using a replication technique, and have to choose among three different possible systems $A$, $B$, and $C$.
The first system $A$ uses the primary backup approach to tolerate $f$ out of $f+1$ faulty components. 
It assumes crash faults and experiences some fail-over time if the primary crashes, so there can be some period of time in which requests cannot be processed. 
Further, it does not degrade in service functionality. 
The second system $B$ uses BFT SMR, employs $n=3f+1$ nodes to tolerate $f$ faulty nodes, and can withstand Byzantine faults. 
A faulty node does not lead to a failover time, because requests are actively processed by all nodes and also service functionality does not degrade. 
The third system, $C$, also uses SMR but assumes only crash faults and thus only requires $2f+1$ nodes. 
If we compare these systems using the different resilience metrics, \textit{degree of replication}, \textit{fault assumption coverage}, \textit{service degradation}, and \textit{failover time}, we will notice that none of the systems is pareto-optimal (see Figure~\ref{fig:resilience_measures}). 
While $A$ offers the lowest replication overhead and can thus tolerate the proportionally largest amount of faulty components ($f$ out of $f+1$), system $B$ is optimal with respect to fault assumption coverage as it can withstand fault classes that neither $A$ or $C$ consider. 
Meanwhile, $C$ lies in a sweet spot, where it is as good as $B$ when it comes to failover time but it can tolerate a proportionally higher amount of faulty components.
However, this comes at the price of making restrictions on what kind of faults can be considered. 
Now, if we want to argue about which of these systems offers the best resilience for some specific service, we need to make trade-off decisions, depending on which resilience metrics are the most important to our specific scenario and environment.

\textbf{Quality of service and degrees of preservation}. Bishop et al. mean by resilience that a system may degrade in its \textit{quality of service}~(QoS) under stress over time, but is able to recover~\cite{bishop2011resilience}. 
Since resilience is associated with requirements of concrete services, the authors argue that QoS metrics can be defined and applied to quantify resilience in respect to an application. 
They explain that there are \textit{degrees of preservation}~\cite{bishop2011resilience} which are basically discrete levels that are used to define how much certain quality aspects of a system can be preserved, e.g., availability and confidentiality. 
These metrics are easier to define for availability than confidentiality, because we can use performance metrics.
However, even for confidentiality and integrity, discrete approaches for quantifying resilience exist, e.g., a QoS measure for confidentiality could be the time needed to brute-force a key of the specific length that is used by the service.
Data integrity can be at least specified binarily: whether it has been violated or not~\cite{bishop2011resilience}, although approaches for using integrity constraints probabilistically exist as well~\cite{khoussainova2006towards}.
In recent works dealing with resilience in networked systems, resilience can be viewed and measured as the integral of maintaining critical functionality of the system, e.g., reflecting the state over time, such as the proportion of functioning system nodes or functionality that is still working~\cite{ganin2016operational,bellini2019cyber}.

\begin{figure}[tb]
	\centering
	\includegraphics[width=0.4\textwidth]{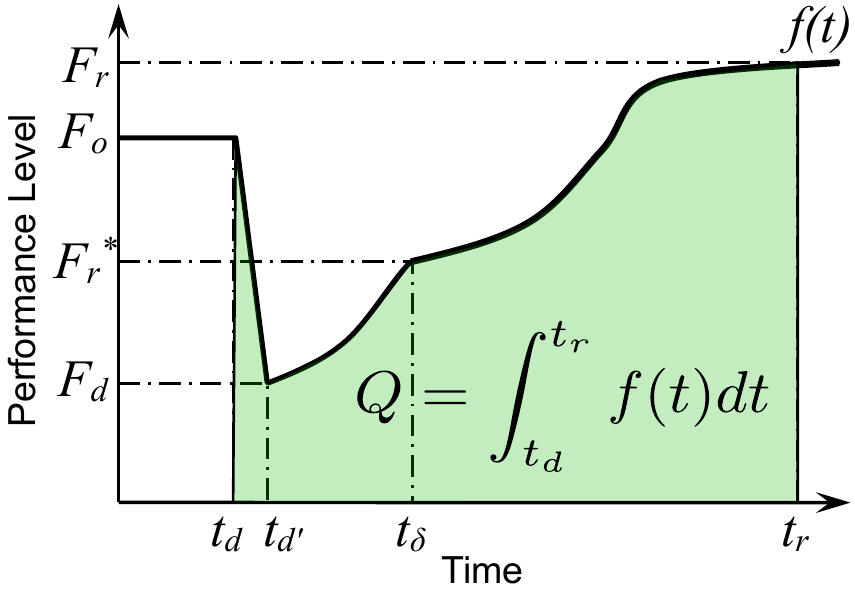}
	\caption{A performance curve shows the relation between time and 
		 functionality during a disruption~\cite{vugrin2011resilience}.}
	\label{fig:performance_curve}
\end{figure}

\textbf{Absorption capacity, adaptive capacity, and recovery capacity.}
Vugrin et al. describe resilience more generally by defining three general concepts of a system as \textit{resilience capacities}, namely the \textit{absorptive capacity}, the \textit{adaptive capacity}, and the \textit{recovery capacity}~\cite{vugrin2011resilience}. 
The 
\textit{absorptive capacity} is the degree to which the system can withstand (thus absorb) disruptions/impacts and prevent or minimize consequences to the system with little effort. It can be viewed as an inherent property of the system. While this is a very general concept, in the context of distributed systems, this resembles the category of \textit{compensation} (or more generally, \textit{error handling}). Further, the \textit{adaptive capacity} means the degree to which a system can self-organize and recover its performance in response to an impact, e.g., by considering internal aspects that manifest over time after a disruption happened. This may require the system to take a lot of internal effort to change (adapt to a disruption) during a recovery period~\cite{vugrin2011resilience}. 
Lastly, the \textit{recovery capacity}
defines the ability of a system to be repaired easily. Figure~\ref{fig:performance_curve} illustrates how these abstract concepts can be quantified: Let $F_d$ be the system functionality in terms of performance that could be preserved after a disruption (which happens at time $t_d$ in the diagram) and $F_0$
the performance level of the initial, unimpaired system. Then, the proportion of  $F_d/F_0$ measures the absorptive capacity of the system while
the proportion $F_r/F_0$ defines the \textit{adaptive capacity} with $F_r$ being the system's performance after it has fully recovered (thus adapted to the disruption).

\textbf{Runtime metrics.} Andersson et al. describe a set of runtime metrics that can be used to quantify system resilience~\cite{Andersson19}: (i) metrics that measure \textit{the continuity of correct service}, e.g., the reliability of a system such as the mean time to failure, (ii) metrics that can quantify the readiness of a correct service to measure the availability (in particular maintainability) of the service, e.g., the ratio of time the system spends in acceptable states with respect to the total observation interval and (iii) metrics that measure the overall, accumulated quality $Q$ of delivered service, which can be leveraged to assess the degradability of the system, e.g., in respect to some performance level which can be defined by the integral of observed performance quality over some time interval (see Figure~\ref{fig:performance_curve}). 

\textbf{Cost metrics: resilience has a price tag.} Making a system resilient is often associated with mechanism-specific costs and limitations. Redundancy-based mechanisms that mask faults usually demand the provision of additional resources (computing components, space, time) and raise costs for system operators, thus making resource-efficiency an important design goal for resilience mechanisms~\cite{kapitza2012cheapbft}. In IoT systems, a new approach could consist of exploiting
the natural redundancy of functionality across devices to compensate for failures while decreasing costs. For example, different devices in a smart home context can report similar events, e.g., a motion sensor and a video camera~\cite{terry16towards}.
 Further, an incorporated resilience mechanism may cause some overhead in the system (e.g., additional computational steps or messaging, use of cryptographic primitives) which can lead to decreased performance during fault-free execution. As an example, in replication protocols, the replicated service is usually benchmarked against the non-replicated service as a baseline to reason about the incurred performance overhead. Metrics that cover these aspects could be motivated by either an economic view (e.g., monetary costs that occur for a system operator) or QoS perspective (impairment of service performance due to resilience mechanism execution).

\subsection{Adjustability}
\label{sect:adjustability}

Given the observations from the previous section, matching specific resilience requirements can have different implications, e.g., with regard to providing additional hardware resources or the selection of resilience mechanisms that might be necessary for a system operator to align their system to satisfy these requirements. We summarize our findings as follows:

\begin{itemize}
\item Fault-model assumptions~(e.g., the \textit{fault-tolerance coverage}~\cite{avizienis04basic}) 
have an impact on the proportion of faulty components that can be tolerated in the system and thus define how much 
redundancy
 (i.e., the \textit{degree of replication}~\cite{budhiraja1993primary}) is needed. For instance, more redundancy is needed to tolerate Byzantine faults compared to crash faults.  
This also means that, given the same number of provided components, we might be able to tolerate more faulty components under the crash-fault model than under the Byzantine fault model.
 \item Fault-model assumptions might not reflect reality accurately enough~\cite{avizienis04basic}. Failures that are assumed to occur independently might actually be correlated, such as when redundant components are attached to the same power supply. This may result in requiring more additional hardware resources than initially planned.
 \item Different views towards how a resilient system should behave might be derived from application-specific requirements. For these, \textit{quality-of-service metrics} may help to define at which degrees of preservation~\cite{bishop2011resilience} a system can operate when under disruption.
  \item Graceful degradation as a resilience mechanism is strongly related to the application and its domain.  
 It is thus virtually impossible to adjust resilience in this regard without adapting the application itself.  
 \item Adjusting security levels might be hard because security is a property that generally cannot easily be quantified.
Still, it can make sense to enforce the use of specific cryptographic primitives or to use adequate key lengths~\cite{bishop2011resilience}.

 \item Resilience properties can be of either qualitative type (a certain quality of service level can be delivered or not, e.g., confidentiality: cryptographic primitives are in place or not) or quantitative (like availability where metrics like uptime or an upper bound of tolerable faulty components exist). This differentiation is not strict. Some ``rather qualitative'' attributes, like confidentiality or integrity, can be viewed quantitatively by measuring the difficulty of breaking underlying cryptographic primitives (e.g., measuring key length to reason about brute force difficulty). Typical quantitative properties like reliability and maintainability come with reasonable metrics that can help system operators to specify desired values.
 
 \item The specification and implementation of desired resilience properties might vary across different IoT architectural layers. For instance, confidentiality requirements might differ between clouds, where demanding AES-256 encryption might be a reasonable choice, and sensor landscape where the use of lightweight block ciphers (with smaller key length specified) is a good trade-off to incorporate security for resource-constrained devices with limited power.
\end{itemize}

%% file: diagrams/radarchart.tex
  \begin{figure}[th]
	\centering
	\begin{tikzpicture}
	\tkzKiviatDiagram[label distance=1cm, scale=0.5,
	radial  = 4,
	gap     = 1,  
	lattice = 3]{
		\begin{tabular}[c]{@{}c@{}} \small{~~~~~~fault} \\ \small{~~~~~~~~assumption} \\ \small{~~~~~~~~coverage} \end{tabular}, 
		\mbox{\small{degree~of~replication}}, 
		\begin{tabular}[c]{@{}c@{}}\small{failover~} \\ \small{time}\end{tabular},
		\mbox{\small{service~degradation}}
	}
	\tkzKiviatLine[ultra thick,color=ourblue,mark=ball,  mark size=4pt, ball color=ourblue,
	fill=ourblue!20,opacity=.5](3,1,3,3)
	\tkzKiviatLine[ultra thick,mark=ball, ball color=ourorange,
	mark size=4pt,color =ourorange, fill=ourorange!20,opacity=.5](2,2,3,3) 
	\tkzKiviatLine[ultra thick,mark=ball, ball color=ourgreen,
	mark size=4pt,color =ourgreen, fill=ourgreen!20,opacity=.5](2,3,2,3)      
	\node[anchor=south west,xshift=20pt,yshift=-22pt] at (current bounding box.east) 
	{
		\begin{tabular}{@{}lp{6cm}@{}}
		\cellcolor{ourgreen} & \small{$A$ Primary-backup approach ($n=f+1$)}\\
		\cellcolor{ourblue} & \small{$B$ BFT SMR ($n=3f+1$)} \\
		\cellcolor{ourorange} & \small{$C$ CFT SMR ($n=2f+1$)}\\
	
		\end{tabular}
	};
	\end{tikzpicture}
	\caption{Radar diagram comparing the resilience of replicated systems using different resilience measures. 
	Increased distance from the center is better, e.g., a failover-time result is further from the center when it is shorter, a service-degradation result when it does not or only marginally degrade service functionality.}
	\label{fig:resilience_measures}
\end{figure}

%% file: sections/mechanisms.tex
\section{Resilience Mechanisms and their Application Requirements}
\label{sect:resilience-mechanisms}

In this section, we analyze and discuss concrete resilience mechanisms from a more practical, implementation-level view, e.g., in respect to the constraints they impose on the application. This discussion includes a broader range of redundancy mechanisms~(Section~\ref{sect:redundancy-mechanisms}), monitoring mechanisms~(Section~\ref{sect:monitoring-mechanisms}), protection mechanisms~(Section~\ref{sect:protection-mechanisms}), and recovery mechanisms~(Section~\ref{sect:recovery-mechanisms}).

 When it comes to resilience in IoT services, we also distinguish different roles. While \textit{users} access the functionality of IoT services and demand their resilience, \textit{administrators} maintain the system. They rely on the autonomous functioning of resilience mechanisms within the system, e.g., when it comes to monitoring components, compensating faulty components, or initiating the recovery of components to preserve system functionality. On the other hand, \textit{developers} create concrete IoT applications (or application components) and while doing so, might need to make some considerations at design-time to support their resilience. 

Assembling and incorporating a set of resilience mechanisms into an IoT platform leads to a design challenge: Every resilience mechanism may have its own \textit{constraints}, such as interfaces that need to be implemented by the application developer,  and also \textit{constraints}, e.g., regarding the execution of the application or its programming model.  
Ideally, a configurable and resilient IoT service execution platform should  ascertain the availability of specific interfaces, as far as they are provided by an application, and automatically derive which resilience mechanisms can be applied.
The constraints specific to a resilience mechanism may complicate (i) the design of an execution platform that can leverage all these constraints, as well as (ii) the porting of existing IoT applications to match this platform.
In this section, we aim to conduct a survey to discuss the constraints of a selection of resilience mechanisms and present an overview of our findings in Table~\ref{tab:resilience-mechanisms}.
 
	\input{tables/resiliencemechanisms.tex}

\subsection{Redundancy Mechanisms}
\label{sect:redundancy-mechanisms}
To make systems more resilient by design, we can anticipate the presence of faults within the system and employ redundancy among hardware, software, or data to overcome possible faults. \textit{Redundancy mechanisms} typically try to mask a specific proportion of faults which are then tolerated and do not lead to a failure. Thus, these mechanisms are a type of \textit{fault tolerance}, in particular \textit{compensation}~\cite{avizienis04basic} or sometimes referred to by the term \textit{absorption}~\cite{vugrin2011resilience}. Further, redundancy mechanisms can be classified depending on how they work e.g., what is being provided (and performed) redundantly such as processing, data storage, or networking.

 \textbf{Auto-scaling.} Scalability describes the capability of a software system to address a growing workload by using more resources. Accordingly, horizontal scalability denotes the capability of a distributed 
 application to increase its performance by increasing the number of resource units~(such as physical servers or virtual machines) available to the application~\cite{seybold20kaa}. Applications that support \emph{Elastic 
 Scaling} support the action of \emph{scaling}~(out or in) while the application is running, hence without downtime and ideally without impacting application performance~\cite{seybold20kaa}. Finally, \emph{auto-scaling}
 provides means to automatically adapt the scale-out factor of an elastically scalable application depending on current or even predicted workload.

 According to~\cite{qu18autoscaling} supporting auto-scaling requires workload monitoring, the existence of an application performance model, a controller matching actual~(or predicted) workload against the model, and 
 deciding on the application's scale-out factor. Moreover, an elastic infrastructure is required to dynamically acquire and release resource units as well as internal or external mechanisms to balance the workload among available resource units. 
 
 Under these circumstances, auto-scaling can be used to increase the resilience of an application by adapting capacity to workload and hence, mitigating possible unavailability due to overload situations. On the downside, naive auto-scaling makes an application vulnerable to denial of service attacks that provoke unnecessary scaling actions and leads to high costs. Auto-scaling has been an active area of research in the cloud era and many different approaches exist focusing on heuristics~\cite{bauer19chameleon,ilyushkin18experimental} and language-level~\cite{achilleos20camel}. The latter approach is also used by auto-scaling features offered by Amazon's AutoScaling\footnote{\url{https://aws.amazon.com/de/autoscaling/}} and Kubernetes\footnote{\url{https://kubernetes.io/docs/tasks/run-application/horizontal-pod-autoscale/}}.
  
\textbf{State-machine Replication.}
\textit{State-machine Replication~(SMR)}~\cite{schneider1990implementing}, also known as active replication, is an approach for tolerating faults by abstracting the service functionality~(i.e., the processing of client requests) as a deterministic \textit{state machine} which is then replicated on multiple independent server replicas. 
These replicas maintain a \textit{consistent state} by applying the same sequence of state transitions.
In practice, this can be achieved by letting replicas deterministically process client requests in the same order.
The requests can be ordered by utilizing a form of \textit{totally-ordered or atomic broadcast} between clients and the replica group.
Typically for resilience, such a broadcast mechanism is implemented by a multi-consensus protocol, whereas in recent years the effectively same functionality has also been provided by blockchains~\cite{vukolic2015quest}. 
Whereas the broadcast can be easily hidden from the application developer, the deterministic execution cannot.
If there are means to convert code unaware of determinism into deterministic code then developers still have to avoid certain operations, library functions, or language concepts as they cannot be converted into a deterministic variant.
An example of such means is DJ, a deterministic run-time for Java~\cite{domaschka2013comprehensive}.
However, mostly such means are not widely available or employed.
Examples for non-deterministic application aspects include \textit{multi-threading}, \textit{random number generation} or any sources of randomness, \textit{time measurements}, \textit{external calls}, or any kind of interaction with some external service. 
Further, some sources of non-determinism might appear unintuitive to the developer, e.g., when it comes to the usage of internal libraries. 
For instance, when iterating over the keys of a Java \texttt{HashMap}, there is no explicit guarantee given on a deterministic order.
Client-replica interactions, like broadcasting a request to the replicas or having the client wait for a specific number of matching responses are also automatically coordinated by the SMR protocol~\cite{schneider1990implementing}. 
Practical Byzantine Fault-Tolerance (PBFT)~\cite{castro1999practical} is a prominent example.
A more recent, still maintained, and feature-rich replication library is BFT-SMaRt~\cite{bessani2014state}. 
In BFT-SMaRt, the replica set is dynamically scalable which is helpful, e.g., to add more replicas at run-time or also to remove replicas if desired (e.g., to re-integrate replicas after completed maintenance). 

A modular architecture helps to separate the library into exchangeable components for different concerns, e.g., \textit{state transfer}, \textit{reconfiguration}, or \textit{consensus}.
The library provides interfaces that can be used by application developers to implement execution behavior, customize state transfer or employ specific client-server interaction models such as publish-subscribe.

\textbf{Passive Replication.}
In passive replication, also called \textit{primary-backup} approach~\cite{budhiraja1993primary}, a single server or system, the \textit{primary}, is responsible for processing client requests and distributes \textit{state updates} to the backups.
Only when the primary fails, a failover occurs, and one of the backup replicas takes over as the new primary.
In order to be able to take over, the new primary has to start from the latest known state and re-execute requests that have occurred since then.
Thus, these requests have to be retained in some way, e.g., by logging to stable storage or by broadcasting to all servers.
The longer the update interval, the more involved the recovery becomes. Additionally, clients using the system have to be made aware of a changing primary, which becomes rather more difficult in the context of IoT, where e.g., resource-constrained devices in the field need to switch to a different edge device. Hokeun et al. present a solution to this problem of migrating IoT devices to fail-over backup systems~\citep{hokeun17sst}.
Very short update intervals need more communication bandwidth than SMR and are usually avoided.
A typical misconception is that passive replication does not require deterministic execution.
If there is multi-threading in the primary or at least some of the request processing produces side effects in other services, the execution must be deterministic, too.
Additionally, the replication system has to retrieve the application state for updates. 
Sometimes, this is not easily possible without involvement of the application developer, as (i)~due to multithreading a consistent state has to be coordinated with request processing, or (ii)~not all state is actually relevant for updates.
In these cases, state has to be compiled by the application.
In order to save communication costs, the replication system may then compute incremental or compressed state updates by preserving previous versions.
More general limitations of this technique are, that---in contrast to state-machine replication---passive replication does not support the Byzantine fault model, and the response time of the passively replicated system to a client request is much higher during recovery time after a primary failure.

A recent example of this is shown in Ambrosia~\cite{goldstein20ambrosia}, which utilizes checkpointing and application logging to resume execution on fail-over secondary servers when a primary fails. Fail-over time, however, was measured to be roughly 3 orders of magnitudes higher than normal case latencies.
CESSNA~\cite{harchol20cessna} presents a solution to the problem of proper checkpointing, request logging, and replay in a distributed edge computing scenario where information about the order and status of requests are dispersed in the edge, as is often the case in IoT systems.

\textbf{Partition-tolerant Data Redundancy.}
By partition-tolerant redundancy we refer to techniques allowing storing data redundantly so that in case of partitions every partition can still access the data.
This would be helpful in an IoT system where the edge gets disconnected from cloud services.
An edge service could redundantly maintain some data that is usually stored in the cloud. 
Edge services synchronize the data with the cloud on best effort. 
In principle, this is very similar to caching because the goal is to benefit from proximity to data for faster access. 
The availability of the edge service increases by preserving functionality of the edge even if the connection to the cloud is disrupted. 
If connectivity between edge and cloud is broken, services running in the edge can still process (important) events by working on their local data. 
Mechanisms that address partition-tolerant data redundancy often work inside architectures where data is being viewed either as a stream or as a state. 
According to the CAP theorem, however, such mechanisms have to sacrifice consistency~\cite{brewer00cap}.

In a \textit{data-as-state} architecture, as soon as data is changed in both partitions, e.g., in edge and cloud, inconsistencies occur and have to be reconciled.
Reconciliation can be application-specific and, in the worst case, even need user interaction.
The usage of CRDTs~\cite{shapiro11crdt} (conflict-free replicated data types) can allow for automatic reconciliation. In this approach, replicas employ the eventually strong consistency model where all replicas eventually converge when no further updates are issued. In IoT systems, CRDTs can be used to provide a reliable storage solution at the edge~\cite{kopestenski2019erlang}.

In \textit{data-as-a-stream} architectures, ideally data flows from producers to consumers along individual processing stages which form
a pipeline. Such a pipeline can be distributed~(and moved) between edge and cloud by
deciding where to place these processing components while data is streamed through the pipeline and buffered at each system site. A recent approach is presented by CEFIoT~\cite{javed2018cefiot}, which describes a crash fault-tolerant architecture that allows placing
processing components 
in both edge and cloud. CEFIoT uses a data transport layer that
employs a pub/sub messaging framework (Apache Kafka), so data streams can be
buffered and replicated across a cluster. Using Apache Kafka, the
data pipeline maintains network fault tolerance by allowing data buffering locally
at the edge during periods in which Internet connectivity is being temporarily disrupted. 
Multiple pub/sub topics are used to buffer data streams, hence making data available for the cluster even during a reconfiguration of individual processing stages. Data is buffered in topics, which are also replicated to other Kafka instances in case some instance becomes unresponsive.


 \textbf{Redundant Network Links.}
Redundancy can also be applied when it comes to networking. While there are protocol-layer--specific solutions to establish reliability over unreliable network links~(such as cyclic redundancy codes or re-transmission of lost packages), the failure of individual network links or network devices can also be tolerated by employing redundancy among these resources. 
For instance, Hasan et al. propose a method for fault-tolerant routing in the IoT between a larger number of interconnected devices, which is implemented by constructing, recovering, and selecting $k$-disjoint multipath routes that guarantee connectivity even after the failure of up to $k - 1$ paths~\cite{Hasan2017optimizing}. \textit{Frontier}~\cite{o2018frontier}, which is a distributed and resilient edge processing
platform for IoT devices, uses network path diversity and selective network-aware replay to recover from transient network failures. 
 Redundancy-based models have also been developed and implemented~\cite{sinche2018assessing}, which allow assessing reliability aspects. Since applications are usually not aware of how reliability in networking is handled, there are no concrete constraints for the application developers.


\subsection{Monitoring and Analysis Mechanisms}
\label{sect:monitoring-mechanisms}




Monitoring and analysis encompass a variety of different mechanisms usable for the detection of changes and harmful conditions of a system. These mechanisms can be divided into different groups like performance, security, and system status, whereas performance monitoring is primarily used for detecting changes in hardware-related performance metrics, e.g., detecting insufficient hardware components that may lead to failures in the availability of services. 
Security monitoring mechanisms usually focus on actively detecting attacks, such as intrusions, based on different detection strategies~\cite{zarpelao2017survey}. 
Furthermore, monitoring and analysis mechanisms can be divided into different phases. These encompass monitoring, which is the actual collection and storage of data, aggregation, which encloses filtering and aggregation of data, and alerting, which analyzes the data and triggers alerts if necessary.    
In general, monitoring systems provide the essential functionality of detecting changes and harmful conditions enabling timely manual or automated intervention.



 
\textbf{System Monitoring.}
System monitoring focuses on retrieving and analyzing system status information. 
Thereby, the monitored objects reach from single components to complete infrastructures. 
Status information is retrieved using various tools, e.g., by utilizing the SNMP protocol or custom system information. 
Their analysis can be rather diverse, reaching from rule-based systems where some thresholds are manually defined (e.g., ``storage medium below 10~Gb of free space'') to more sophisticated ones like anomaly detection using machine learning techniques. 
Due to the focus on system status information these monitoring mechanisms do not impose any constraints on the application. 
However, the runtime environment, operating systems, or physical components  have to provide such system information.
As an example, DARPA~\cite{ibrahim16} proposes the integration of cheap, reliable read-only hardware clocks into IoT devices, to enable a secure heartbeat protocol for monitoring an IoT network, in order to mitigate attacks on individual devices.

\textbf{Intrusion Detection Systems.}
Intrusion detection systems (IDS) are a widely evaluated field of research with early efforts dating back to the 1980s~\cite{anderson1980computer}. IDS are designed to detect malicious actions, called intrusions, within an infrastructure, and to report those either to a human or to other services capable of taking appropriate action, hence, ensuring resilience~\cite{zarpelao2017survey}. In order to detect such intrusions there exist a multitude of different approaches starting with the simplest one, called specification based, where rules of a specific behavior of a system are manually specified and every violation of one of those rules is categorized as an intrusion. This method is very time-consuming and error-prone, since missing a rule or changing the system's behavior always requires human input. A second approach is to monitor if the behavior of a system matches an attack signature previously recorded and stored in a database. Effort-wise this approach is less time demanding than the previous, since the signature only has to be captured once. However, there is a delayed response time when the system is confronted with new attacks. Anomaly based systems are coping with this challenge by first observing the normal behavior of the monitored system and afterwards detecting deviations of this normal behavior which are then classified as intrusions. The main advantage of this technique is its self-adaption to changing normal behavior and the automatic detection of new attacks. As an example, Parker et al.~\cite{parker19} implemented an anomaly detection strategy utilizing deep extraction and mutual information selection, which increases the interpretability of machine learning models and is optimized for resource-constrained and dynamic IoT environments.

However, the IoT with its different infrastructure in comparison to traditional ones, holds new challenges for IDSs. On the one hand, IoT devices are usually restrained regarding their computational power and with that partly incapable of executing IDS services. On the other hand, there is a multitude of heterogeneous devices, gateways, etc. increasing the number of, e.g., placement strategies of IDS systems and the variety of used communication protocols~\cite{zarpelao2017survey}. There are currently several research studies  to improve the status quo in this field~\cite{anthi2019supervised,raza2013svelte,breitenbacher19,ali19}, like Breitenbacher et al.~\cite{breitenbacher19}, which present a host-based IDS with low performance overhead, thereby extending the deployment capabilities of IDS on IoT devices. The authors utilize a whitelist approach to identify malicious processes running on the IoT devices themselves. Other mechanisms include a behavior monitor for blockchains running on IoT devices classifying their behavior regarding maliciousness~\cite{ali19}. 
A further category of the attack surface prominent in IoT infrastructures is physical attacks, which open up new strategies for the detection of such attacks. An example in this category would once again be the work of Ibrahim et al.~\cite{ibrahim16}, which proposed a distributed heartbeat protocol capable of detecting the absence of devices. In this case, the protocol assumes an attack and initiates mitigation strategies. Other approaches observing the physical domain of IoT devices are proposed by Yasaei et al.~\cite{yasaei20}, which detect anomalies in sensor data and, based on that, predict possible malicious actions.
However, many questions remain to be answered~\cite{zarpelao2017survey}.

\textbf{Introspection.}
IDS systems have often been placed directly in a host, e.g., using host-based agents for monitoring, or they have been placed on routers to monitor the network traffic. However, network-based IDS can not capture details about what exactly is going on within a host. 
Introspection techniques monitor virtual machines or containers at \textit{run-time and from the outside}. For instance, \textit{virtual machine introspection} (VMI) is a technique that can be employed as the basis to build an intrusion detection architecture on top of it~\cite{garfinkel2003virtual}. This gives some distinct advantages: The IDS can achieve an excellent view of the host system while being isolated and thus more protected from manipulations than in-host agents (if the host becomes compromised). In fact, the monitored virtual machine is not even aware that it is being monitored. Within the context of IoT systems, the applicability of VMI is a bit narrowed: it is particularly applicable in the cloud layer of an IoT infrastructure. There is a range of already existing VMI tools~\cite{jain2014sok} that are suitable for monitoring a VM for possible intrusions, such as the detection of root kits. VMI works application-agnostic and does not impose concrete constraints on an application (except running in a VM environment).

\textbf{Honeypots.}
Honeypots aim to trick attackers into thinking they are interacting with a real system instead of a purposely set up mockup. The purpose is often to either study the attacking behavior and patterns, or to distract attackers from attacking real systems. Recent research work~\cite{pa2016iotpot, luo2017iotcandyjar, guarnizo2017siphon, dowling2017zigbee} discusses how the use of honeypots in the IoT landscape (e.g., mimicking IoT devices) can identify novel IoT-related threats and attack patterns. As an example, Tanabe et al.~\cite{tanabe20} identified the common IoT botnet infrastructure and its behavior. Their results showed that IoT botnets, while being less sophisticated in their architecture compared to traditional infrastructure botnets, are disposable, meaning they are only used once. This dynamic nature allows them to be resistant against countermeasures like blacklisting.
To integrate honeypots into an IoT execution environment and deploy those automatically, the concept of shadow honeypots is very promising. Shadow honeypots exist alongside ``real'' IoT components/applications and mock these by instantiating the same component/application in a secured environment. Afterwards, intruders either actively access those shadow honeypots or get redirected to them once they are found out~\cite{anagnostakis2010}.

\textbf{Plausibility Checks.}
Plausibility checks are a very important instrument in detecting faulty and even malicious components~\cite{kuemper2018valid,rumez2018security}. Here, the reported values of, e.g., a monitoring component are compared to either the known environment boundaries of the monitored service or values of other monitoring components or both, to check whether the reported values are plausible. 
For instance, based on knowledge we can tell that a temperature sensor located in a living area reporting 200°C is most likely wrong. 
Apart from this simple example there are also more sophisticated methods, like machine learning value prediction and anomaly detection, ensuring the plausibility of values. In respect to IoT built around the concept of sensors measuring data these checks can be an essential part of ensuring resilience. 
Checks can be performed on different architectural layers, such as cloud or edge layer, depending on the quality of data and requirements of application-specific information. However, many of these checks can only be realized with application-specific knowledge, meaning that only the application logic itself can decide whether a given value is plausible or not. Hence, the application developer needs to implement this resilience mechanism and, simultaneously, appropriate interfaces for reporting the processed data.




\subsection{Protection Mechanisms}
\label{sect:protection-mechanisms}
 A variety of mechanisms are especially designed to shield a system from external and possibly malicious harm, to prevent (malicious) faults from occurring. We call these \textit{protection mechanisms}.

\textbf{Encryption.}
Encryption is a security mechanism that aims at ensuring the confidentiality of data while in transit, at rest, or in processing to avoid that data is disclosed to some unauthorized entity, e.g., a malicious eavesdropper.
As IoT systems consist of heterogeneous
device classes that come with different computational resources, a variety of cryptographic primitives and protocols are utilized to match application-specific requirements. Lightweight cryptographic primitives~\cite{singh2017advanced,yang2017survey} should be considered to be used with \textit{power-constrained} devices, e.g., lightweight block ciphers might come with smaller block and key sizes than commonly employed. As for asymmetric cryptography, elliptic curve cryptography~(ECC)~\cite{miller1985use} is a better fit than, e.g., the RSA~\cite{rivest1978method} algorithm, because RSA employs rather large key sizes.
For encryption schemes, the overall goal in IoT is finding a practical trade-off: ensure a reasonable degree of security with the least amount of overhead while saving computational power, battery life, and bandwidth since they are sparse resources. As an example, Ferretti et al.~\cite{ferretti19} introduced a symmetric proxy re-encryption scheme ensuring encryption among IoT devices and edge computing instances even in case of a failed cloud connection. Another approach is presented by Shi et al.~\cite{shi16}, which introduces a new ultra-light weight encryption scheme capable of protecting confidentiality of data of white-box attack scenarios. They implemented this scheme so that the main memory requirement decreases to a fraction of comparable approaches, enabling its utilization within the IoT.
Similarly, Gu and Potkonjak~\cite{gu2018multipuf} presented a secure multistage physically unclonable function (PUF) design which allows for key distribution, key storage and rekeying for purposes of IoT device authentication with ultra low power consumption characteristics.
From an application-level view, the process of encryption is mostly transparent, so that encryption schemes can be easily implemented and applied on a broad variety of devices. However key management can become a challenge in larger sized systems.

\textbf{Signatures.}
Generally, signatures can guarantee essential security properties, in particular, integrity, data origin authentication, and non-repudiation. They can be applied e.g., to verify data and application integrity.
Often, signatures are used as authentication devices on messages. In the field of IoT, a security goal could be to protect the integrity of sensor data. It has been shown that integrity protection, e.g., with the use of ECC signatures is in principle possible nowadays even on constrained devices but with some restrictions~\cite{pohls2015json, bauer2016ecdsa, mossinger2016towards}.

\textbf{Verification.}
Verification has the ambition of guaranteeing that an application satisfies some specified properties, e.g., by generating correctness proofs or finding counterexamples. This allows the detection and removal of internal development faults within the system during its design phase. 
In the IoT, software bugs can have implications on the safety of systems as unintended interactions within the physical spaces may lead to unsafe or insecure environments~\cite{celik2019verifying}, which motivates code verification.
Verification efforts have contributed to building secure, safe, and dependable systems: for instance, model checking based approaches~\cite{nguyen2018iotsan, celik2018soteria} exist that allow for detecting interaction-level flaws between IoT components possibly causing unsafe and dangerous
physical states and can also help to validate safety properties. Other approaches suggest the use of code-level analyses by symbolic execution~\cite{tabrizi19design} in contrast to model checking for its faster analysis, better detection rate, and fewer false positives.
In the same vein, approaches like CASCA~\cite{delledonne2018casca} aim to improve the process of designing IoT hardware by creating domain-specific hardware modeling languages which can reason about e.g., side-channel attack-resilience in the circuit design phase of IoT silicon. 
Gatouillat and Badr suggest using linear logic to automatically compose applications of components following their smart object model~\cite{gatouillat17verifiable}. Particularly, a smart object formally describes an artifact structure and behavior.
Also, a learning approach for model-based testing of communication in the IoT, such as implementations of different MQTT brokers~\cite{tappler2017model} has been proposed, which can identify differences in various implementations that adhere to the same MQTT specification (hence indicating possible bugs), using models that are automatically learned from the implementations.

\textbf{Privacy Filters \& Privacy Preserving Techniques.} Due to the nature of IoT and cyber-physical systems, a lot of sensitive data of users (e.g., related to location, health, or daily activities) can be generated and aggregated, forwarded, and processed between IoT architecture layers. This raises the demand for techniques to preserve privacy, e.g., by enforcing specific policies to protect the privacy of users of IoT applications (which can also be implemented in a network-centric approach~\cite{sivaraman2015network}). An approach could implement local differential privacy obfuscation to leverage IoT data analytics at the edge, so that data is aggregated and distilled without disclosing users’ sensitive data before sending it to the cloud~\cite{xu2018distilling}.
Other approaches ensure privacy by applying blockchains and using smart contracts to control and enforce connection rights~\cite{loukil21data}.

\textbf{Identity Management, Authentication and Authorization Services.}
The scale, growth and heterogeneity of IoT systems also impact the manageability of identities leading to novel challenges when designing authentication and authorization services. Those new challenges encompass not only the external attackers on the cloud and edge layer, but also on the sensor layer, e.g., by placing malicious sensors among the regular ones with the goal of connecting them to the infrastructure. Since IoT extends into the physical world, system components might be deployed in adverse environments.
Generally, the goal of identity management is to protect the system against an attacker forging identities (spoofing attacks) or spawning fake identities (Sybil attacks). Identities are often used with authentication and authorization services to prevent an attacker from accessing information or consuming resources unauthorized.
 OnboardICNg is an example of a protocol that prevents malicious IoT devices from joining a trusted information-centric network (ICN),
and prevents mislead honest devices to become part of a malicious ICN~\cite{compagno16onboardICNg}. 
Because IoT devices may collect sensitive data or impact the safety of their environment, a reasonable choice is to realize the principle of least privilege. Recent research work has shown that embedding and process awareness
of IoT devices impose a natural sequencing of accesses, which can
be used to enforce history-based access control policies~\cite{tandon18hcap}.
 Further, identity management as well as authentication and authorization services are often implemented by middleware and protocols below the application level, so that they can be used with little interference to the application logic and do not have to be considered as a distinct problem by an application developer~\cite{hernandez2015toward}.
The IoT introduces new challenges: the availability of authentication and authorization services deployed on edge computers needs to be preserved even under, e.g., network failures or attacks. Recent research work proposes a resilient authentication and authorization framework for IoT which allows an IoT device to securely migrate to another trusted edge computer if its own local authorization service becomes unavailable (e.g., due to a Denial-of-Service attack)~\cite{kim20resilient}. A further approach relies on controlling the flow of information for event-based IoT systems at the brokers to restrict which devices may communicate with each other~\cite{fuentes19brokering}. The consideration of \textit{home-limited channels}~\cite{ji20authenticating} that
can be accessed only within a house yet remain inaccessible for attackers outside the house also provides an interesting application for authenticating devices in a smart home~\cite{ji20authenticating}.
 Lastly, securely pairing new IoT devices can be a problem in many public IoT systems -- such as in hospitals, factories or laboratories -- where a multitude of different devices provide critical services and share private data in unsafe environments with potentially many close-by attackers. This, as well as the lack of sophisticated input options on many small IoT devices, can preclude the usage of usual pairing techniques based on, e.g., proximity or entering pairing codes. The works of Li et. al present solutions for these scenarios~(\cite{li2019touch},~\cite{li2020t2pair}). They utilize sensing and matching of semi-random user input sequences like petting a device or using universal input sensors like buttons or knobs to securely pair devices in the presence of multiple malicious attackers.

\textbf{Sensor Fusion.}
Sensor Fusion describes the mechanism of fusing the values of different measurement points (sensors) in order to get results of higher quality or better plausibility~\cite{terry16towards}. There are various methods for reaching these goals, starting with simple aggregation of data from homogeneous sensors and calculating the median value, for example. A more elaborate approach is to aggregate data from heterogeneous sources to get a better impression of an environment, hence preventing possible faults on one sensor by comparing to the environment. Another idea is to interpolate or predict possible values of single sensors by using information of surrounding sensors, thus checking future values for correctness. Linked to these approaches, there are also sensor data quality analyses trying to quantify the quality of received data~\cite{kuemper2018valid}. This mechanism, however, is highly application specific, meaning that the application developer is in charge of implementing it.


\subsection{Recovery Mechanisms}
\label{sect:recovery-mechanisms}

Another pillar of resilient system design is anticipating faults and errors that occur, and making the system capable of detecting and recovering from them. 
\textit{Recovery mechanisms} aim to steer a system back to its intended, functional state. 
Unlike redundancy mechanisms which mask present errors and faults, recovery  
 transforms a system state that contains errors and faults into a new system state in which errors do not exist and faults will preferably not activate again~\cite{avizienis04basic}.

\textbf{Recovery: Checkpointing, Rollbacks, and Roll-forwards.}
Since unpredictable errors can occur within system components, it might be reasonable to anticipate this and provide resilience mechanisms that, once a failure is detected, can restart a component and recover its application state to continue execution. 
A basic approach is to let applications periodically create checkpoints~\cite{koo1987checkpointing, tong89low} which might require the provision of interfaces for acquiring and persisting state and for restoring the application from that state.
The latter is typically called a rollback.
In rarer cases, e.g., with some sort of transactions involved, application state could also be rolled forward to a new and error-free state.
One example is the completion of NTFS transactions according to log entries belonging to a particular transaction~\cite{Nagar97}.
Depending on the application, checkpoints can become large.
Solutions to create smaller entities are incremental checkpointing or the deployment of techniques like event sourcing, which record state changes at a fine granularity level.
Ambrosia~\cite{goldstein20ambrosia} is a recent example of a system that provides resilience by logging requests (and periodically performing checkpoints) to replicated storage prior to execution and guaranteeing correct
state reconstruction during replay-based recovery. Its key features include high performance, support for non-determinism as well as support for language and machine heterogeneity, which also makes it a suitable choice for IoT applications.
In IoT systems, applications might be deployed across heterogeneous devices which also involve resource-constrained devices. A recent approach works at design time to automate an optimized rollback-recovery for constrained scenarios by employing a co-design of firmware, runtime and compiler transformations to create multiple resilient variants of an application which can reduce the checkpointing overhead~\cite{automating2018}. 

Compared to rollbacks, roll-forwards also purge detected errors from the system state but without retrying from the last correct checkpoint.
For instance, in a modular redundant system (e.g., Duplex system), a roll-forward lets the execution of tasks continue while the fault diagnosis and recovery functions are performed concurrently (by an external spare component). In particular, on both the correct and the faulty component the task execution continues \textit{forward}, beyond the last checkpoint where disagreement occurred~\cite{pradhan1994roll}. After identifying which component is faulty by employing the spare component parallel to continuing task execution on both of the other components, the state of the correct component can be copied to the faulty one, hence transforming its erroneous state to a state without errors, from which it can continue execution.

In event-driven IoT applications, an approach for data repair consists of recording events and replaying computations that depend on corrupted data. \textit{SANS-SOUCI}~\cite{lin2019data} is a recent example that combines a functions-as-a-service architecture for the
event-driven system, append-only data structures to persist
application state durably, and distributed, causal dependency tracking for efficient replay of dependent computations across
multi-tiered (device, edge, cloud) IoT deployments.

\textbf{Graceful Degradation and Upgrade.}
\textit{Graceful degradation} means that an IoT system, in case of faults, does not entirely fail but tries to uphold as much functionality as possible.
Graceful degradation is a recovery mechanism that enters the scene when faults can no longer be tolerated by other resilience mechanisms. 
For instance, the loss of connectivity between cloud and edge may leave the edge without access to resource-consuming and enabling services provided by the cloud layer.
In this case, a degradation mechanism could switch to less resource-consuming alternatives in the edge by sacrificing some of the original quality, e.g., precision, accuracy, efficiency, throughput, or consistency.
The degraded service could even abstain from executing certain tasks and providing specific functionality at all.
Thus, the degradation can affect \textit{function}, \textit{quality}, \textit{performance}, or any combination thereof.
Graceful degradation approaches are well known for the design and operation of distributed embedded systems~\cite{nace01familiyDegradataion, knight04survivability, glass09degradation}, but have recently also been introduced for IoT scenarios, e.g., for a video surveillance application~\cite{chang2014bringing}, a smart-office case study~\cite{willocx2019qos}, and a drone application~\cite{yoon17virtualdrone}.
Degradation often depends on defining a set of different \textit{service levels} that determine the graduations of quality of service (QoS) a system can operate at, where different levels might require a different set of functioning IoT components.
Further, degradation as a resilience mechanism is application specific, as only the application domain can define service levels and interpret them by deploying different implementation alternatives.

IoT is not restricted to a certain application domain as it can reach from private home environments to even globally distributed networks.
Also, degradation decisions depend on faults that no longer can be tolerated by other resilience mechanisms. 
Thus, there is a need to combine system support with application-specific degradation mechanisms. 
Application developers should be guided on how to define service levels and how to integrate degradation with system-provided support mechanisms, e.g., fault-tolerance means, monitoring of failure events, etc. 
One approach is to use a product-family approach to define service levels~\cite{nace18degradation}.
Besides, the switch between service levels should also be supported by framework or system functions in order to reduce tangled application-, service-level- and system-specific code fragments in IoT applications.

A degraded system may automatically switch back to the originally desired service level, or at least to a better level, if the cause of the degradation has disappeared or changed.
This procedure is called \emph{graceful upgrade}.
Ideally, an automatic upgrade procedure exists, e.g., the system monitors the availability and functionality of needed components and re-decides about its next service level.
If not, a system will need manual reconfiguration to redeploy at the original service level.

\subsection{Combined Mechanisms}

So far, we have analyzed and discussed a broad range of resilience mechanisms. In practice, these mechanisms can and will be combined in one way or the other. For instance, several of the mechanisms listed inherently depend on other mechanisms. Further, the architectural characteristics of IoT systems imply that different mechanisms should be used on different layers of the architecture. 

\textbf{Inherently combined mechanisms.} 
Several mechanisms rely on other mechanisms on a lower level. First and foremost self-monitoring plays a crucial role for many mechanisms. The fact that the system has some degree of self-awareness allows it to detect when to apply countermeasures. Similarly, monitoring is necessary to identify bottlenecks in the system, detect failures and attacks, and thus creates a decision basis for other mechanisms to activate.
Recovery is a fundamental building block enabling individual IoT components to join back into the system. It is therefore the basis of a more general and automated procedure of handling failures. For instance, primary-backup replication applies rollback or roll-forward when a backup replica takes over the primary role.
Finally, replication protocols aiming at Byzantine fault tolerance usually apply signatures, while identity management, authentication, and authorization services make use of encryption.

\textbf{Combinations for enhanced resilience.} 
Mechanisms can also be combined to enhance the overall resilience. This includes for instance a mutual support of IDS and honeypots as presented by Baykara et al.~\cite{baykara2018} and anomaly detection mechanisms integrated with shadow honeypots~\cite{anagnostakis2010} both aiming at the reduction of false positives. The first approach focuses on zero-day attack detection of signature-based IDS, where honeypots enable the autonomous signature generation of such attacks. Otherwise those types of IDS would not be able to detect zero-day attacks. The second combination enhances the detection accuracy of the single mechanisms, by having the honeypot verifying positive tagged network traffic by the anomaly detection mechanism.

Moreover, a broad variety of mechanisms should be considered at design time to help withstanding and absorbing changes. These include code validation mechanisms, but also protection mechanisms like cryptographic primitives. The latter help shielding a system from attacks while redundancy mechanisms can be used to compensate for a specific number of faulty components in the system.

\textbf{Control-loops and self-healing.} 
The paradigm of self-healing has been designed in the early 2000s~\cite{kephart2003vision} and are representative of systems utilizing different mechanisms to achieve reconfiguration with the aim of satisfying their requirements in a changed environment or possibly degrade gracefully. These mechanisms always include some form of continuous monitoring, essential for detecting unforeseen changes that might hinder the system's resilience, in combination with mechanisms capable of executing appropriate counteractions that can maintain the requirement satisfaction, e.g., graceful degradation. 
The general procedure can be described by a MAPE loop~\cite{kephart2003vision,tsigkanos2019towards}: Constant (M)onitoring of the environment and reflecting changes in a model, (A)nalyzing the model for requirement violations, (P)lanning appropriate countermeasures, and finally (E)xecuting those and updating the model. A MAPE-K loop enhances MAPE with a knowledge base~\cite{arcaini15modeling,rutten17mapek}, while MAPE-SAC introduces security aspects in the control loop~\cite{jahan2020mapesac}. Auto-scaling approaches are usually realized via a MAPE(-K) loop.

Tsigkanos et al. see self-adaptation capabilities as a core feature of future autonomous resilient IoT systems~\cite{tsigkanos2019towards}. Muccini et al. stress that IoT systems may require more than one control loop and that different control loops may be overlapping~\cite{muccini18selfadaptive}.
Ozeer et al. present a failure management protocol for stateful IoT applications in a dynamic environment~\cite{resilienceFog2018}. It applies a control loop combining recovery techniques~(using checkpointing) and monitoring as well as failure notification and reconfiguration.
Current research in this area includes the work of Seeger et al. who introduce an approach for failure detection and mitigation strategies of IoT (edge) devices for complex software tasks~\cite{seeger20}. The mitigation strategies include an optimal task allocation strategy for distributed task execution on IoT devices. Dias et al. introduce a pattern language for specifying self-healing strategies for IoT systems~\cite{dias20pattern}.
A major challenge that has not received much attention is the design of a resilient monitoring system and control loop respectively able to take decisions from incomplete information and being able to deal with failures in the monitoring or execution chain.

\textbf{Architectural combinations.}
Due to their size and geographical distribution as well as the large attack surface, IoT systems are prone to failures even more than traditional distributed systems. Consequently, resilience mechanisms need to be applied on all layers of an IoT architecture. Here, a particular challenge for enabling resilient IoT systems their massive heterogeneity. Network capacities and latency in different layers may vary by six and more orders of magnitude. Similar holds for the computational and storage capacities. 
These differences and the fact that not each device is alike imposes constraints on the use of resilience mechanisms and the orchestration of compute tasks. For instance, mechanisms such as state-machine replication are only ill-suited for the sensor and IoT edge layer due to the rather weak network connectivity, large network latency, and limited computational capabilities. On the other hand, the use of wireless communication technology inherently creates redundancy provided that multiple gateway nodes are placed in the range of a sensor. Hence, the installation of an IoT system limits resilience possibilities. This dependency on the actual geo-distribution of devices furthermore requires that mechanisms such as graceful degradation need to be tailored individually to each IoT system.

%% file: tables/resiliencemechanisms.tex
\footnotesize
\begin{tabularx}{\textwidth}{|s|b|j|k|}

	\hline 
	\emph{\begin{hyphenrules}{nohyphenation}Resilience Mechanisms\end{hyphenrules}} & \emph{Resilience Goal} & \emph{Fault Model} & \emph{Constraints for the Application} \\  \hhline{|=|=|=|=|}

	\rowcolor{gray!15} \multicolumn{4}{|c|}{redundancy mechanisms} \\* \hline
	
	\emph{auto-scaling} & adapt capacity to workload & overload & application needs to provide metrics as decision basis, architecture needs to be scalable \\ \hline
	
	\emph{state-machine replication} & tolerate up to f faulty replicas out of n replicas ($n > 2f$ for crash faults) & Byzantine faults, crash faults  & deterministic service execution, interfaces for state transfer, execution and client-server interaction model \\ \hline

	\emph{primary-backup replication} & tolerate up to f faulty replicas out of n replicas ($n > f$ for crash faults) & crash faults & support for state updates and request logging \\ 
	\hline 
	
		\emph{partition-tolerant redundancy } & tolerate connectivity loss to cloud, preserve functionality in edge & network connectivity loss between edge and cloud & interfaces to define data to be cached in edge, fallback functions \\ 
\hline 
\emph{data redundancy} & tolerate faulty nodes & up to $f$ faulty nodes & has to deal with inconsistencies \\ 
\hline 
\emph{redundant network links} & tolerate failure of network links & up to $f$ broken links & --- (often handled by underlying protocols or middleware) \\ 
\hline 
	
	\rowcolor{gray!15} \multicolumn{4}{|c|}{mechanisms related to monitoring} \\* \hline	

	\emph{monitoring} & general surveillance and anomaly detection of component behavior & malicious intrusions and faults & component provides interface to report its own health status to the outside world \\ 
	\hline 
		\emph{intrusion detection systems} & detection of intrusions for mitigation purposes & malicious intrusions & application provides support for plausibility checks on data \\ 
	\hline 
	
		\emph{introspection} & monitor a virtual machine or container state dynamically from outside, assert specific policies & malicious alterations & application needs to run in some VM or container \\ 
	\hline 
		\emph{honeypots} &  analysing attacking patterns and strategies; distracting attackers &  malicious intruders & if honeypots for components (e.g., sensors) should be automatically deployable, application needs to provide a description for execution platform\\ 
	\hline

	\emph{plausibility checks} & detection of implausible sensor data, hence, prevention of failures in service execution & malicious modifications & either integrated in some IDS (anomaly detection) or handled by application itself (application-specific) \\ 
	\hline
	
		\rowcolor{gray!15} \multicolumn{4}{|c|}{protection mechanisms} \\* \hline	
	
	\emph{encryption} & protect the confidentiality of data in transit at rest or while processing & malicious eavesdroppers & key and identity management, support for standard protocols like TLS \\ 
	\hline
		\emph{signatures} & protect the integrity of data & malicious modifications & key and identity management, support for standard protocols like TLS \\ 
	\hline 
			\emph{verification} & detect and remove software bugs. assert correctness of  applications  & development faults & e.g. creating a model for model-checking an application (along with its interactions)\\ 
	\hline 
	\emph{privacy filters} & protect the privacy of users & unwanted leakage of personal information  &  domain-specific knowledge on which and how data needs to be masked or aggregated\\ 
	\hline
		\emph{identity management, authentication and authorization services} & protect against forging or spoofing identities and/or attackers trying to access resources without permission  & Sybil / spoofing attackers,  unauthorized resource consumption &  --- (often handled by underlying protocols or middleware)  \\ 
	\hline 
	\emph{sensor fusion} & prevention of distribution of faulty sensor data & malicious intruders & handled by application itself due to application-specific knowledge requirements  \\  
	\hline 

	\rowcolor{gray!15} \multicolumn{4}{|c|}{recovery mechanisms} \\* \hline

	\emph{rollback} & periodically perform checkpoints and if necessary, e.g., restart some component and restore state from checkpoint   & unpredictable errors occurring in the system & provide interface for checkpointing and for restoring application state from checkpoints \\ 
\hline
	\emph{rollforward} & periodically perform checkpoints and upon error detection perform fault diagnosis \textit{concurrently}; transfrom  a faulty component's state into a new state without errors   & unpredictable errors occurring in the system & provide interface for checkpointing and for restoring application state from checkpoints \\ 
\hline
	\emph{graceful degradation} & continue service delivery on best efforts basis, e.g., under a restricted service level  & \multirow{2}{=}{faults 
		 that can not be fully tolerated and need to be circumvented}  & \multirow{2}{=}{define QoS requirements, specify service levels and application-specific behavior as well as transitions between service levels}\\
\cline{1-2}
	\emph{graceful upgrade} & recover original service level after graceful degradation & & \\
\hline

		\caption{Overview over application requirements of resilience mechanisms.}
	\label{tab:resilience-mechanisms}
\end{tabularx} 

\normalsize

%% file: sections/related-work.tex
 \section{Related Work}
 \label{sect:related-work}
Various research studies and analyzes the current status of incorporating resilience in IoT and distrusted systems, thus presenting a broad variety of emerging challenges and solutions in dependability and security~\cite{tsigkanos2019towards, terry16towards, FTIoT2019, Abreu2017resilient, Gia2015fault, Zhou2015supporting, Hasan2017optimizing, Christidis16blockchains}.

\textbf{Related systematization of knowledge papers.}  To begin with, the extensive taxonomy of Avi{\v{z}}ienis et al. for dependable and secure computing is helpful for establishing a common understanding of key concepts in the broader field of distributed systems~\cite{avizienis04basic}. 
Ratasich et al. present a roadmap towards a more resilient IoT for cyber-physical systems~\cite{resilienceIoTRoadmap2019}. They focus on summarizing presented techniques, e.g., on fault tolerance, anomaly detection, or self-healing, and outline present and future challenges 
for dependability and security in such systems. 
 Moghaddam et al. present a systematic mapping study in which they identify and categorize a set of methods for achieving fault tolerance in the IoT~\cite{FTIoT2019}. The authors focus more on a statistical comparison of how current research employs existing techniques, e.g., for which 
  different architectural styles these are employed, which attributes are aimed to be improved, and which research trends are emerging. 
The work of Tsigkanos et al. proposes a research roadmap for resilient IoT system design and future challenges~\cite{tsigkanos2019towards}. The authors identify high-level resilience mechanisms and the need to deal with disruptions, and sketch future directions for 
engineering resilient IoT systems. For instance, they propose having the edge infrastructure consumed as a full-fledged utility, abstracting business logic management from the infrastructure, or autonomous control and self-healing capabilities.
Welsh et al. survey resilience techniques for cloud environments~\cite{cloudResilienceSurvey20}. The authors first analyze the state-of-the-art of techniques for different cloud components and subsequently discuss current limitations and challenges. Further, Sequeiros et al.
survey existing attack and system modeling techniques that are applicable to IoT and cloud computing~\cite{sequeiros20attack}.

\textbf{IoT research trends with resilience focus.} Terry et al. discuss fault-tolerance techniques, such as \textit{compensation}, in the context of IoT to make IoT systems more resilient~\cite{terry16towards}. The authors propose to seek novel and lightweight solutions.
For instance, devices like sensors or actuators are often fail-stop, and hence
application-level fault tolerance can be provided without employing state-machine replication but instead with simple fail-overs to additional, correct components. An interesting trick for IoT systems could be to leverage pre-existing redundancy, e.g., by letting IoT devices
discover and select nearby devices that can report similar events, such
as motion or presence, or that support similar actions, like turning on a light. Additionally, devices could automatically connect to nearby operating hubs and then switch hubs as soon as failures occur~\cite{terry16towards}. This can be utilized to reduce replication costs. Further, concepts such as \textit{virtual services}~\cite{Zhou2015supporting} that consolidate data from more than one physical sensor can also be considered for resilient design in service-oriented IoT architectures.
The use of blockchains and smart contracts to build
reliable IoT ecosystems is discussed in recent works~\cite{Christidis16blockchains, fernandez2018review, durand2018resilient, han2018evaluating}.
Blockchains can provide distributed, trust-less, and resilient service execution and can thus be employed, e.g., as a billing layer for marketplaces of services between devices.
Abreu et al. propose a modular IoT middleware for smart cities that comprises a distinct component called \textit{Resilience Manager}, which is in charge of supervising activities by accessing a \textit{Monitor} module~\cite{Abreu2017resilient}. Additionally, it employs a \textit{Protection and Recovery} module that can perform actions upon detection of faults, e.g., relying on other modules such as topology control as well as placement and migration modules.
 Gia et al. study resilience in the domain of \textit{e-Health} systems and present an approach for increasing reliability on top of a 6LoWPAN communication infrastructure~\cite{Gia2015fault}. The authors use backup routing between nodes and advanced service mechanisms to maintain connectivity in case of failing connections between system components.
Mart{\'\i}n et al. introduce a multi-level platform architecture for resilient IoT applications~ \cite{martin2019edge}. Their platform employs containerization of components, abstracts functionality of devices behind so-called shadow devices, and achieves fault tolerance through the replication of physical devices and automatic reconfigurations. The authors show their approach can be practically used, e.g., in the field of health monitoring. Javed et al. propose CEFIoT as a fault-tolerant architecture for the IoT~\cite{javed2018cefiot}. CEFIoT employs the Apache Kafka platform for data replication in both edge and cloud and uses Kubernetes for fault-tolerant management and automatic reconfiguration of the processing pipeline in case of, e.g., connectivity failures. Costa et al. present a voting mechanism for edge computing that lets the edge network validate computations that were performed in a multi-cloud environment~\cite{costa2019dependable}.

\textbf{Resilience engineering and resilience techniques.}  Hukerikar et al. present a catalog of resilience design patterns~\cite{hukerikar2017resilience}, which is aligned to the field of high performance computing but is also applicable to other computer engineering domains since reoccurring problems and their solutions are described on a very abstract level. Also, their work is highly didactic and tries to guide computer scientists to incorporating resilience-enhancing ideas into concrete system designs. Jackson et al. present a survey on a set of abstract, top-level resilience principles for engineered systems, such as \textit{absorption}, physical and functional \textit{redundancy}, and \textit{layered defense}~\cite{jackson2013resilience}.
Dias et al. present a pattern-language for self-healing IoT systems~\cite{dias20pattern}. In their work, the authors describe a catalog of patterns (building blocks), which can be efficiently combined to craft  self-healing systems. These patterns can be divided into two main classes: \textit{~Error detection} and \textit{recovery and maintenance of health}. They also explain how different health pattern actions can be sequenced to achieve service restorations towards normal state, thus enabling resilience for a system.

%% file: sections/conclusion.tex
\section{Conclusions}
 \label{sect:conclusion}

Resilience is often being referred to in different contexts and with different ambitions in mind. Generally, resilience has the meaning to preserve different system properties such as availability, confidentiality, integrity, reliability, maintainability, or safety \textit{when the system encounters change}. Colloquially speaking, a system is being expected to ``withstand'' change or ``bounce back'' to a functional state. ``Change'' can be further differentiated, e.g., into planned changes or unplanned disruptions, such as internal or external faults or attacks. On this note, resilience comprises not only dependability but also security, i.e., should be able to deal with attacks.

The IoT  has unique architectural and technical properties with specific challenges such as scalability, evolvability, maintainability, heterogeneity, and complexity. In particular, we need to consider that system architectures span over several IoT layers, and resilience mechanisms should therefore also work across layers. Further, we may need to consider that there are multiple administrative domains and diverse views to the system. Resilience requirements may differ between the views of, e.g., an application user, developer, system operator, or IoT hardware provider. As IoT systems change, these requirements might evolve over time, too. 

Quantifying resilience is a multi-dimensional approach, which depends on the goals that are considered. Concepts to measure the effectiveness of resilience mechanisms vary. For instance, quality-of-service metrics, fault-tolerance coverage, or cost metrics exist, but have to be carefully chosen to best represent an application's resilience constraints. Further, it is crucial to enable IoT systems' self-adaptation capabilities, since resilience does not only mean ``tolerance'', but often also an approach to ``trying to uphold service under best efforts''.

A broad variety of applicable resilience mechanisms exists. We divided them into redundancy, monitoring, protection, and recovery techniques. The goal of these mechanisms is to make a system capable of compensating faults (redundancy), detecting faults or attacks (monitoring), shielding against attacks, preventing the occurrence of harm (protection), and steering the system back to a well-defined functional state (recovery). The incorporation and combination of a well-chosen set of such resilience mechanisms allow an IoT system to self-heal and thus become more resilient, but mechanism-specific constraints need to be respected and considered by application developers at design time and by the execution platform at runtime.